\pdfoutput=1
\documentclass[11pt]{article}
\usepackage[ruled, vlined, linesnumbered, noresetcount]{algorithm2e}
\usepackage{times}
\usepackage{epsfig}
\usepackage{amsthm}
\usepackage{amsmath}
\usepackage{amsfonts}
\usepackage{amssymb}
\usepackage{enumerate}
\usepackage{graphics}
\usepackage{color}
\usepackage{mathtools}
\usepackage{url}

\SetKwFor{ForAll}{forall}{do}{end forall}

\usepackage{graphicx}
\usepackage[numbers]{natbib}
\setcitestyle{square}

\date{}

\SetKwFor{ForAll}{forall}{do}{end forall}

\newtheorem{theorem}{Theorem}

\newtheorem{lemma}{Lemma}
\newtheorem{prop}{Proposition}
\newtheorem{definition}{Definition}

\newtheorem{claim}{Claim}

\newtheorem{fact}{Fact}

\DeclareMathOperator*{\argmax}{arg\,max}

\SetKwFor{ForAll}{forall}{do}{end forall}

\newcommand{\remove}[1]{}

\newcommand{\Active}{{{\sf D}}}
\newcommand{\A }{\Active}

\def \k {k}
\def \l{\ell}
\def\tw{{\tt{tw}}}
\def\nd{{\tt{nd}}}

\def \f {f}
\def \h {h}
\def\bh{{\bf h}}
\def\bbf{{\bf f}}

\def \Y {{Y}}

\def \r {{r}}
\def \s {{s}}
\def \q {{q}}
\def \Ecd {{E_{cd}}}
\def \Vc {{V_c}}
\def \Vd {{V_d}}
\def \Lc {{L_c}}
\def \Ld {{L_d}}
\def \Lcn {{\Lc\mbox{-neg}}}
\def \Lcp {{\Lc\mbox{-pos}}}
\def \Lcg {{\Lc\mbox{-guard}}}
\def \ecdo {{e_0^{cd}}}

\def \ecdm {{e_\s^{cd}}}
\def \ecdj {{e_j^{cd}}}
\def \vco {{v_0^c}}
\def \vcu {{v_1^c}}
\def \vcr{{v_r^c}}
\def \vci {{v_i^c}}
\def \vch {{v_h^c}}
\def \Lcd {{L_{cd}}}
\def \Lcdn {{\Lcd\mbox{-neg}}}
\def \Lcdp {{\Lcd\mbox{-pos}}}
\def \Lcdg {{\Lcd\mbox{-guard}}}
\def \Mcd {{M_{cd}}}
\def \Mcdn {{\Mcd\mbox{-neg}}}
\def \Mcdp {{\Mcd\mbox{-pos}}}
\def \Mcdg {{\Mcd\mbox{-guard}}}
\def \Iccd {{I_{c:cd}}}
\def \Idcd {{I_{d:cd}}}
\def \Iccdp {{\Iccd\mbox{-pos}}}
\def \Iccdn {{\Iccd\mbox{-neg}}}
\def \Iccdg {{\Iccd\mbox{-guard}}}
\def \Idcdp {{\Idcd\mbox{-pos}}}
\def \Idcdn {{\Idcd\mbox{-neg}}}
\def \Idcdg {{\Idcd\mbox{-guard}}}
\def \MQ {{\sc MQ}}
\def \bug{bag}

\begin{document}

\title{Parameterized Complexity of Immunization in the Threshold Model}

\author{Gennaro Cordasco\\Department of Psychology,\\ University of Campania ``L.Vanvitelli'', Italy 
\and
Luisa Gargano\\ Department of Computer Science,\\ University of Salerno, Italy
\and 
Adele Anna Rescigno \\ Department of Computer Science,\\ University of Salerno, Italy}

\maketitle 
\begin{abstract}
We consider the  problem of controlling the  spread of harmful items   in networks, such as  the contagion proliferation  of diseases or  the diffusion of fake news.
We assume  the linear threshold model of diffusion where each node has a threshold that measures  the node resistance to the contagion. We study the parameterized complexity of the  problem: Given a network, a set of initially contaminated nodes, and two integers $k$ and $\l$, is it  possible to limit the diffusion to at most $k$ other nodes of the network by immunizing at most $\l$ nodes?
We consider several parameters associated to the input, including: the bounds $k$ and $\ell$, the maximum node degree $\Delta$,  the treewidth,  and the neighborhood diversity of the network.
We first give $W[1]$ or $W[2]$-hardness results for each of the considered parameters. 
Then we give fixed-parameter algorithms for  some parameter combinations.\\       

\end{abstract}

\section{Introduction}
The problem of controlling the  spread of harmful items   in networks, such as  the contagion proliferation  of diseases or  the diffusion of fake news, has recently attracted much interest from the research community.
The goal is to try to limit as much as possible the spreading process by adopting immunization measures.
One such a measure consists in intervening on the network topology either blocking some links so that they cannot contribute to the diffusion process \cite{Kimuraetal} or by immunizing some nodes \cite{ER19}.
In this paper we focus on the second strategy: Limit the spread to a small region of the network by immunizing a bounded number of nodes in the network.
We study  the  problem in the linear threshold model where each node has a threshold,  measuring the node resistance to the diffusion \cite{Kempe}. A node gets influenced/contaminated if it receives the item from a number  of neighbors at least equal to its threshold.
The diffusion proceeds in rounds:  Initially only a subset of nodes  has the item and is contaminated. At each round the set  of  contaminated nodes  is augmented with  each node  that has a number of already contaminated neighbors at least equal to its threshold. 

In the presence of an immunization campaign,  the \emph{immunization} operation on a node inhibits the contamination  of the node itself.  Thus, given a network  and   a subset of its nodes, called \emph{spreader} set, that has the malicious item to be diffused to the other nodes in the network, at each round  the set of contaminated  nodes is augmented only with the  nodes  for which the  number of already contaminated  neighbors is at least equal to the node  threshold.

Under this diffusion model, we  perform a
broad parameterized complexity study of the following problem:
 {\em Given a network, a spreader set, and two integers $k$ and $\l$, is it  possible to limit the diffusion to at most $k$ other nodes of the network by immunizing at most $\l$ nodes?
 }

\subsection{Influence diffusion: Related Work}

During the past decade the study of spreading processes in complex networks have   experienced a particular surge of interest across many research areas from  viral marketing, to social media, to population  epidemics. 
Several studies have focused on the problem of finding a small set of individuals who, given the item to be diffused, allow its diffusion to a vast portion of the network, by using the links among individuals in the network to transmit the item itself to their contacts \cite{book}.
Threshold models are widely adopted by sociologists to describe collective behaviours \cite{granovetter} and 
their  use  to study of the propagation of innovations through a network  was first considered 
 in \cite{Kempe}.
The linear threshold model has then been  widely used in the literature to study the problem of influence maximization, which aims at identifying a small subset of nodes that can maximize the influence diffusion \cite{Ben-Zwi,CGMRV,CGRV,asonam,DKT16,Kempe}.

Recently, some attention has  been devoted to the important issue of developing strategies for reducing the spread of negative things through a network.
In particular several studies considered the problem of what structural changes can be made to the network topology in order to  block negative diffusion processes.
Contamination  minimization in linear threshold model by blocking some links has been studied in \cite{cuttingedge,Kimuraetal}.
Strategies for reducing the spread size by immunizing/removing nodes has  been considered in several paper.
As an example  \cite{barabasi,newman}  consider a greedy heuristic that immunize  nodes in decreasing order of out-degree.

When  all the node thresholds are  1,  the immunization can be obtained by a (multi)cut of the network. Some papers dealing with this problem are \cite{Kempe-vaccino,Kempe-cut,bodlander} in case of edge cuts and \cite{Fomin} in case of node cuts.

\subsection{Parameterized Complexity}

Parameterized complexity is a refinement to classical complexity theory in which one takes into account not
only the  input size, but also other aspects of the problem given by a parameter $p$. 
We recall that a problem  with input size $n$ and  parameter $p$ is called {\em fixed parameter tractable (FPT)} if it can be solved in time $f(p) \cdot n^c$, where $f$ is a computable function only depending on $p$ and $c$ is a constant.

We study the parameterized complexity of the studied problem, formally defined in Section \ref{section1}.
We consider several parameters associated to the input: the bounds $k$ and $\ell$, the number $\zeta$ related to  initially contaminated nodes, and some parameters of the underlying network: The maximum degree $\Delta$,  the treewidth  \tw\ \cite{R86},  and the neighborhood diversity \nd\ \cite{L}.  
The two last parameters, formally defined   in Sections \ref{secLBTW} and \ref{secLBND} respectively, are two incomparable parameters of a graph that can be viewed as representing  sparse and dense graphs respectively  \cite{L}; they received   much attention in the literature
{\cite{ALM+,CBFGR,Ben-Zwi,CGRV,itp,CDP,DKT16,FGKKK,GKK18,G,GR,Kn19}}.

\subsection{Road Map}

  In Section \ref{section1}, we formally define  the studied immunization  problem and summarize our findings.
  In Section   \ref{hard},   we give  hardness results for  the considered parameters. In  Section   \ref{Algo},
 we give fixed-parameter algorithms for  some parameter combinations. 

\section{Problem statement}\label{section1}
Denote by $G=(V,E,t)$ a undirected graph where $V$ is the nodes set, $E$ is the set of edges, and $t:V\to\mathbb{N}$ is  a  node threshold function.
We use $n$ and $m$ to denote the number of nodes and edges in the  graph, respectively. 
The degree of a node $v$ is denoted by $d_G(v)$. The neighborhood of  $v$ is denoted by $\Gamma_G(v)=\{u\in V | (u,v)\in E\}$.
In general, the neighborhood of a set $V'\subseteq V$ is denoted by 
$\Gamma_G(V')=\{u\in V |  (u,v)\in E, \ v\in V',\ u\notin V'\}$.
The graph induced by a
node set $V'$
in $G$ is denoted $G[V']=(V', E', t')$ where $E'= \{(u, v) : u, v \in  V', \ (u,v)\in E\}$ and $t'(v)=t(v)$ for each $v\in V'$.

Given the network and a  spreader set $S$,  after one diffusion round,  the influenced nodes are   all those which are influenced  by the nodes in  $S$, that is, have a number of neighbors in $S$ at least equal to their threshold.
Noticing  that nodes in $S$ are already contaminated and cannot be immunized, we can then  model   the diffusion process  as in a  graph which represents the network except  the spreader set. Namely, we consider the graph 
$G=(V,E,t)$ where: $V$ is the set of nodes of the network excluding those in the spreader set, $E\subseteq V\times V$ is the  edge set, and 
$t$ is  the     threshold function $t:V\to\mathbb{N}$ with $t(v)$  equal to the original 
 threshold of the node $v$ in the network decreased by  the number of its neighbors  in $S$.
\begin{figure}[tb!]
	\begin{center}
	   \includegraphics[width=0.8\linewidth]{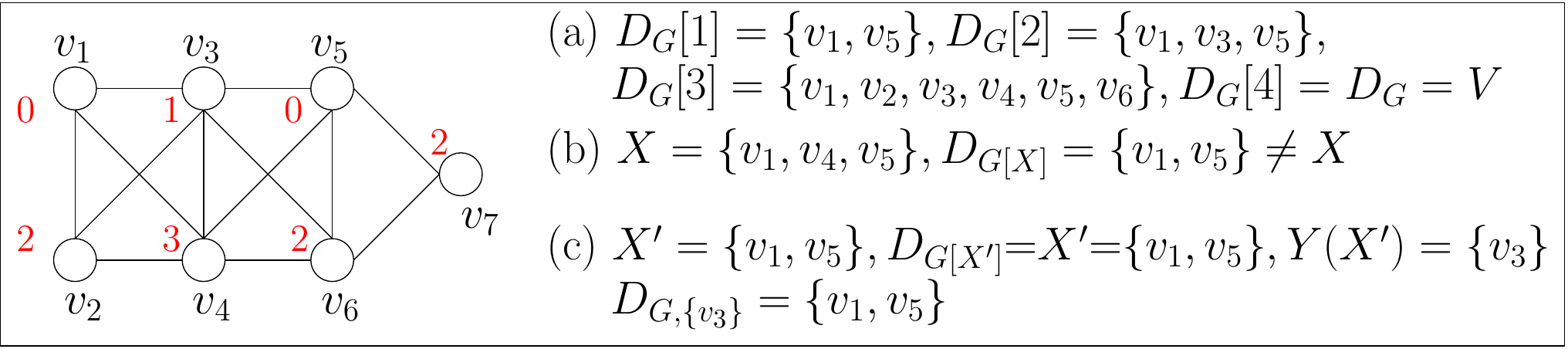} 
		\caption{\small A graph $G$ (node thresholds appear in red). (a) The diffusion process in $G$. (b) An example of $X$ whose $G[X]$ includes nodes not influeced. (c) An  example of immunizing set $Y(X')=\{v_3\},$ which enables to confine the diffusion to $X'=\{v_1,v_5\}$.  
	  \label{Example}}
		\end{center}
\end{figure}

 \begin{definition}\label{active}
 The  {\em diffusion process in $G=(V,E,t)$} in the presence of a set $Y\subseteq V$ of immunized nodes
is a sequence  of node subsets
$\Active_{G,Y}[1] \subseteq 
\ldots\subseteq \Active_{G,Y}[\tau] \subseteq \ldots \subseteq V$\
with

-- $\Active_{G,Y}[1] = \{u | u\in V-Y,\ t(u)=0\}$, and 

-- $\Active_{G,Y}[\tau] = \Active_{G,Y}[\tau-1]\cup \Big\{u  | u\in V-Y,\ \big|\Gamma_G(u)\cap \Active_{G}[\tau-1]\big|\geq t(u)\Big\}$.
\\
The process ends at $\tau^*$ such that  $\Active_{G,Y}[\tau^*]=\Active_{G}[\tau^*+1]$. We set 
$\Active_{G,Y} = \Active_{G,Y}[\tau^*].$
\end{definition}
 
We omit the subscript $Y$ when no node is immunized, that is,  $\Active_G=\Active_{G,\emptyset}$. Moreover,  we assume that for the input graph it holds
$\Active_{G} = V;$
indeed, we could otherwise  remove all the nodes that cannot be influenced, since they are irrelevant to the immunization problem.
In particular,  each   remaining node  $v\in V$ has  $t(v)\leq d_G(v)$, otherwise it could not be influenced. An example is given in Fig.\ref{Example} (a).
\def \IIB {{\sc IIB}}
We are now ready to formally define our problem.
\begin{quote}
    {\textbf {\textsc{Influence-Immunization  Bounding (\IIB)}}}:
Given a graph $G=(V,E,t)$  and  bounds   $\k$ and $\l$,
is there a set $Y$ such that $|Y|\leq \ell$ and $|\Active_{G,Y}|\leq k$?
\end{quote}
For a given set $Y$ we are partitioning the nodes into three subsets: 
The  set $\Active_{G,Y}$ which contains the nodes that get influenced, 
the {\em immunizing}    set  $Y$, 
which has the property  that, if all its nodes are immunized then 
the  diffusion process is 	circumscribed to   $\Active_{G,Y}$,
and  the set $V-Y-\Active_{G,Y}$ of the nodes that, by  immunizing   $Y$,  are not influenced.
\\
We will refer to the nodes in the above subsets as {\em influenced, immunized} and {\em safe}, respectively.

In some cases it will be easier to deal with a different formulation of  {\sc IIB} that starts from the set of nodes  to which  one  wants to confine the diffusion.
Given a set $X\subseteq V$, we define the {\em immunizing  set} $Y(X)$  of $X$  as the set  
 that contains  all the nodes in $V-X$ that  can be influenced in one round by those in $\Active_{G[X]}$, that is, the nodes that get influenced in $X$ when $X$ is isolated from the rest of the graph, namely
 \begin{equation}\label{Y(X)} 
    Y(X)=\{ u   |   u\in V-X, \ |\Gamma_{G}(u)\cap \Active_{G[X]} | \geq t(u)  \}.
\end{equation}
By the above definitions, we have

\begin{equation}\label{GX}
\Active_{G[X]}=\Active_{G,Y(X)}=\Active_{G[V-Y(X)]}\subseteq X;
\end{equation}
 
\noindent
 hence, the  influenced, immunized and safe node sets are  $\Active_{G[X]}$,   $Y(X)$,  $V-Y(X)-\Active_{G[X]}$. 
 \\
For some $X$,  some nodes in $G[X]$ may be not influenced, even though they would in the whole graph $G$ (see Fig.\ref{Example} (b)). However,  it is easy to see that  for each $X$ the set  $X'=\Active_{G[X]}\subseteq X $ is such that 
 $\Active_{G[X']}=X'$  and 
 $Y(X')=\{ u  |   u\in V-X', \ |\Gamma_{G}(u)\cap \Active_{G[X']} | \geq t(u)  \}
 =Y(X)$.
In the following, we will refer as {\em minimal} to a set  $X$ such that $\Active_{G[X]}=X$ (see Fig.\ref{Example} (c)).
\begin{fact}
{\bf ({\sc\bf IIB} equivalent
)}
  $\langle G,k,  \l\rangle$ is a {\sc yes} instance 
  iff there is  a minimal  $X\subseteq V$ s.t.

  \begin{equation}\label{equiv}
  |X|=|\Active_{G[X]}| \leq k \mbox{ and } |Y(X)| \leq \l. \quad 
\end{equation}
\end{fact}

\noindent
\subsection{Summary of results} In this paper we  prove that {\scshape Influence-Immunization Bounding} is:
\begin{itemize}
\item[i)] W[1]-hard with respect to any of the  parameters   $k$,  $\tw$ or $\nd$
\item[ii)] W[2]-hard with respect to the pairs   ($\ell$, $\Delta$), or  $(\ell, \zeta)$;
\item[iii)] FPT  with respect to any  of the pairs 
 $(k,\ell), (k, \zeta), (k,\tw), (\Delta,\tw), (k,\nd), (\ell,\nd)$,
\end{itemize}
 where $\tw$ and $\nd$ denote the tree width and the neighborhood diversity of the input graph and $\zeta=|\{v| v\in V, t(v)=0\}|$ is the number of nodes with threshold 0.
\def \IIB {{\sc IIB}}
\section{Hardness}\label{hard}
In this section we give $W[1]$ or $W[2]$ hardness results for  the considered parameters.
\subsection{Parameter $k$} \label{k}

\begin{theorem}\label{th-k}
\IIB\ is $W[1]$-hard with respect to $k$.
\end{theorem}
\proof 
We give a reduction from    the {\scshape cutting at most $k$ vertices with terminal (CVT-$k$)}  problem studied in \cite{Fomin}: {\em Given a graph $H=(V(H),E(H))$,  $s\in V(H)$, and two integers $k$ and $\ell$,  is there  a set $X_H\subseteq V(H)$ such that $s\in X_H$, $|X_H|\leq k$, and $|\Gamma_{H}(X_H)|\leq \l$?}

To this aim,  construct the instance $\langle G,k-1,\ell\rangle$ of \IIB\, where   $G=H[V(H)-\{s\}]$
and 
$t(v)=0$ for each node $v\in\Gamma_H(s)$ and $t(v)=1$ for each node $v\in V(H)-\{s\}-\Gamma_H(s)$.

Suppose 
$\langle G,k-1,\ell\rangle$ admits a solution. By 
(\ref{equiv}), there exists a minimal set $X$ such that $|X|=|\Active_{G[X]}|\leq k-1$ and $|Y(X)|\leq \ell$.
Noticing that $\Gamma_H(s)\subseteq X\cup Y(X)$, one gets that for $X_H=X\cup \{s\}$ it holds 
$\Gamma_H(X\cup \{s\})=Y(X)$. Hence 
 $X_H=X\cup \{s\}$ satisfies the inequalities $|X_H|\leq k$  and $|\Gamma_{H}(X_H)|\leq \l$ and
 is a solution to {\scshape CVT-$k$}.

Suppose now $X_H=X\cup \{s\}$ is a minimum size solution to {\scshape CVT-$k$}. Then $H[X_H]$ is connected, otherwise the connected component containing $s$ would be a smaller solution. Recalling that in $G$ all thresholds are at most 1, we have that all the nodes in the connected component of a node with threshold 0 get influenced. Hence,

\begin{eqnarray*}
Y(X)&=& \{ u   | \ u \in V-X, \ |\Gamma_{G}(u)\cap \Active_{G[X]} | \geq t(u)  \}\\
&=& \{ u   |   \ u \in V-X, \  t(u)=0  \}\cup \{ u   |  \ u \in V-X,   |\Gamma_{G}(u)\cap X | \geq 1 \}\\
&=&{ \Gamma_H(\{s\}\cup{X})}.
\end{eqnarray*}

\noindent
 As a consequence,
  $X$ is a solution to  \IIB. 
The theorem follows, since Theorem 3  in \cite{Fomin} proves that the latter problem  is $W[1]$-hard   whit respect to $k$.
\qed 

The same reduction, recalling that   Theorem 5 in \cite{Fomin} proves that {\scshape CVT-$k$} is $W[1]$-hard    with respect to   $\ell$,   also gives that 
\IIB\ is $W[1]$-hard with respect to $\l$; however, a stronger result is given in the next section.

\subsection{Parameters $\zeta$ and $\ell$ }
\begin{theorem}\label{zero}
\IIB\  is $W[2]$-hard  with respect to the pair of parameters $\zeta$,  the number of nodes with threshold 0, and $\ell$.
 \end{theorem}
\proof
We give a  reduction from   {\sc Hitting Set (HS)}, which is $W[2]$-complete in the size of the hitting set:
{\em  Given  a collection $\{S_1,\ldots, S_m\}$ of subsets  of a  set $A=\{a_1,\ldots,a_n\}$
and an integer $h>0$, is there a set $H \subseteq A$ such that $H \cap S_i \neq \emptyset,$  for each\footnote{For a positive integer $a$, we use $[a]$ to denote the set of the first $a$ integers, that is $[a] = \{1, 2, \ldots, a\}$.} $i \in[m]$ and $|H| \leq h$?}

 Given an instance  $\langle\{S_1,\ldots, S_m\}, A=\{a_1,\ldots,a_n\}, h\rangle$ of {\sc HS}, we construct an instance $\langle G, n+1,h \rangle$ of   \IIB. 
The graph $G=(V,E,t)$ 
has node set
$$V=I\cup A\cup S,$$
where  $I=\{v_0,\ldots, v_h\}$ is a set of $h+1$ independent nodes,  $A=\{a_1,\ldots,a_n\}$ is  the ground set, and   $S=\{s_1,\ldots,s_m\}$ (each $s_j$ represents the set $S_j$),
  edge set  
$$E=\{(v_i,a_j)\   |\ v_i\in I, \ a_j\in A\}\cup 
\{(a_j, s_t)\ |\ a_j\in A,\ s_t\in S, \ a_j\in S_t\}, $$
and threshold function defined by 
$$t(v)=\begin{cases}
  { 0} & {\mbox{ if $v\in I$}}\\
   { 1 }& {\mbox{ if $v\in A$}}\\
    {{|S_t|=d_G(s_t)}} & {\mbox{ if $v=s_t\in S$.}}
  \end{cases}
$$
Trivially,   $\Active_G[1]=I$,  $\Active_G[2]=I\cup A$, and $\Active_G[3]=I\cup A\cup S=V$.
We prove now  that $\langle\{S_1,\ldots, S_m\}, A, h\rangle$ is a {\sc yes} instance of HS iff
$\langle G,n+1,h\rangle$ is a {\sc yes} instance of \IIB.

Suppose first there exists $H \subseteq A$ such that $|H| \leq h$ and  $H \cap S_t\neq \emptyset$,  for each $t\in [m]$. 
If we consider in $G$ the set of nodes $\tilde Y\subseteq A$ corresponding to the elements of $H$ then 
each node $s_t\in S$  is connected with a node  in $\tilde Y$.  
Consequently, if all the nodes in $\tilde Y$ are immunized, then the  number of influenced neighbors of $s_t$ cannot   reach its threshold $t(s_t)=d_G(s_t)$. Hence, no node in $S$ can get influenced. 
Let then $Y$  be the set obtained by padding $\tilde Y$  with 
nodes in $A-\tilde Y$, so to have
$|Y|=h$.
Clearly, $\Active_{G,Y}=I\cup (A-Y)$ with $|\Active_{G,Y}|=  n+1$.

Assume now there exists  a solution $Y$ of \IIB.
We notice that:
\begin{itemize}
\item[a)] $I\subseteq \Active_{G,Y}\cup Y$ (having all the nodes in $I$ threshold 0, they are  immunized or influenced);
\item[b)] If there exists $v_i\in I\cap Y$, we can update $Y$ to $Y'=Y\cup \{a\}- \{v_i\}$, for any $a \in A-Y$ \\
(this implies that   $\Active_{G,Y'} \subseteq \Active_{G,Y}\cup \{v_i\}- \{a\}$).
\item[c)] If there exists    $s_{t}\in S\cap Y$  
we can update  $Y$ to $Y'{=}Y\cup \{a\}{-} \{s_t\}$,   for  any  $a\in A\cap S_t$\\
(this implies that   $\Active_{G,Y'} \subseteq \Active_{G,Y}- \{a\}$).
\end{itemize}
Using a) and iterating b) and c), we can assume that $Y$ consists of  at most  $h$ nodes in $A$. 
As a consequence  $I\cup (A-Y)\subseteq \Active_{G,Y}$.
If we assumed that $S\cap \Active_{G,Y}\neq \emptyset$,  then   
we would have 
$|\Active_{G,Y}|\geq |I|+|A-Y|+ |S\cap \Active_{G,Y}|> h+1+(n-|Y|) \geq n+1.$
Being $S\cap \Active_{G,Y}= \emptyset$ implies  each node in  $ S$ has some  neighbor  in $Y$. Hence, the set $H$ of  elements corresponding to the $h$ nodes in $Y$ satisfies $H \cap S_t \neq \emptyset,$  for each $t\in [m]$.
 \qed

\subsection{Parameters $\Delta$ and $\ell$}
\begin{figure}[tb!]
	\centering
	   \includegraphics[height=3.1truecm]{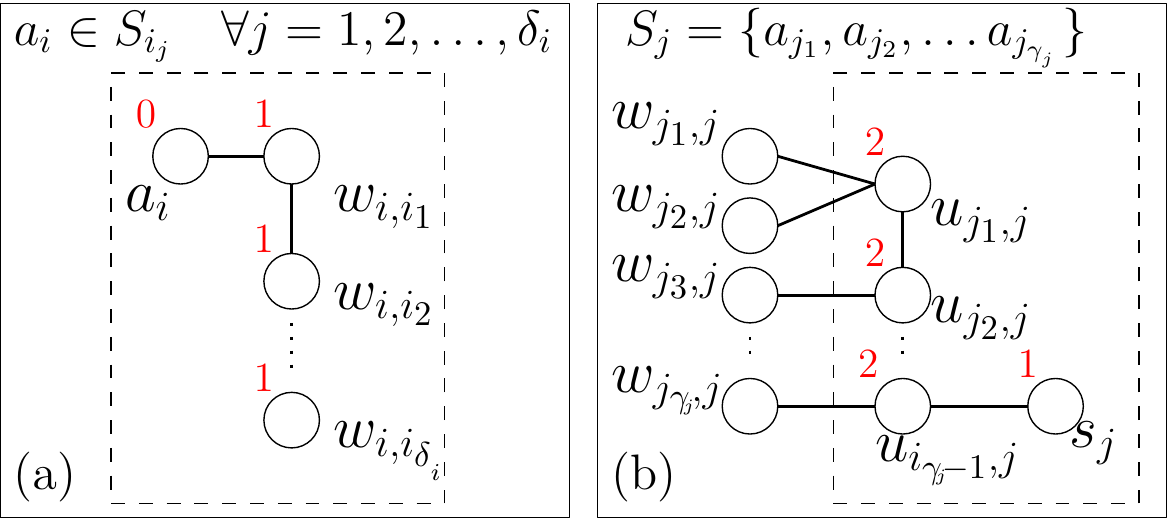}
	   \includegraphics[height=3.1truecm]{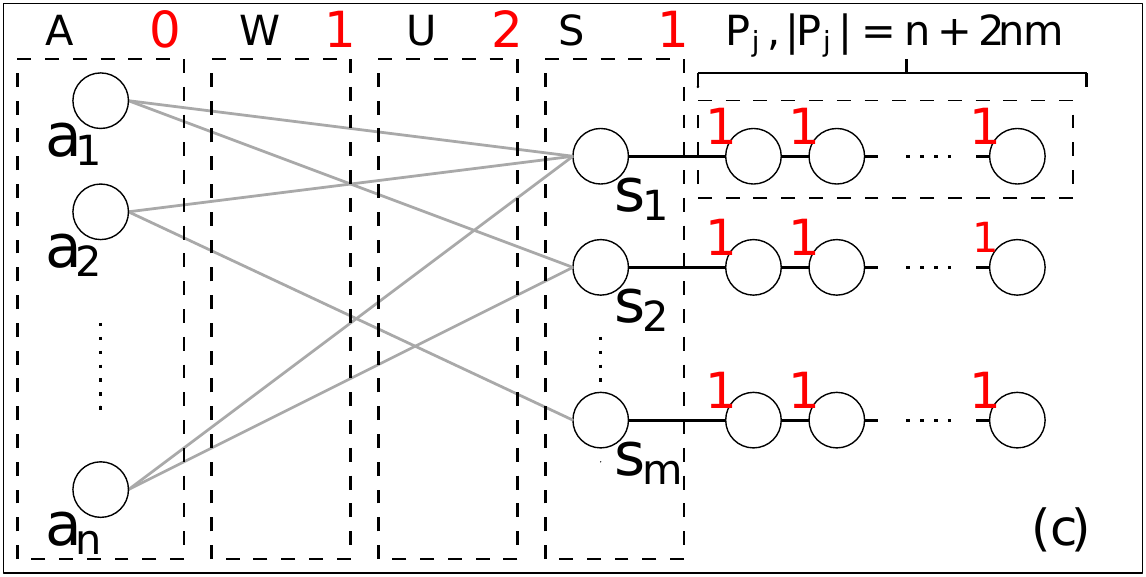}
		\caption{\small (a) The \textit{expansion} gadget. (b) The \textit{reduction} gadget. (c) The graph $G$. 
		\label{GraphDeltaEll}  
		}
\end{figure}

\begin{theorem}\label{deltaEll}
\IIB\  is $W[2]$-hard  with respect to the pair of parameters $\Delta$,  the maximum node degree,
and $\ell$.
\end{theorem}
 Given an instance $\langle\{S_1,\ldots, S_m\},  A=\{a_1,\ldots,a_n\}, h\rangle$  of  HS, we construct an instance $\langle G, k,\ell \rangle$ of   \IIB,
where the maximum node degree is $3$. 
We start the construction of  $G$ by inserting the nodes in $A  \cup W \cup U \cup S$ where   $A=\{a_1,\ldots,a_n\}$ is  the ground set and   $S=\{s_1,\ldots,s_m\}$ (each $s_j$ represents the set $S_j$), while $W$ and $U$ are two auxiliary sets, of at most $nm$ nodes each, that will be used to keep the degree bounded and, at the same time, simulating a complete bipartite connection between $A$ and $S$.
We then add the following \textit{expansion}, \textit{reduction} and \textit{path} gadgets.\\
\\
{\bf Expansion gadgets.} 
 For each $i\in[n]$,  if  the sets containing $a_i$ are exactly  $S_{i_1},S_{i_2},\ldots,S_{i_{\delta_i}}$ then we encode this relationships
 with a gadget,  which includes  four new nodes for each   
$s_{i_j}$, for $j\in[\delta_i]$.
 Namely, we add $\delta_i$  nodes  $\{w_{i,i_1},w_{i,i_2},\ldots w_{i,i_{\delta_i}}\}$
  and the edges 
 $(a_i, w_{i,i_1})$ and   $(w_{i,i_j},w_{i,i_{j+1}})$ for $j\in[\delta_i-1].$\\
\\
{\bf Reduction gadgets.} 
 For each $j\in[m]$,  if $S_{j}=\{a_{j_1},a_{j_2}, \ldots, a_{j_{\gamma_j}}\}$ then we encode this relationships
 with a gadget.  Namely,  we add $\gamma_j-1$  nodes  $\{u_{j_1,j},u_{j_2,j},\ldots, u_{j_{\gamma_j-1},j}\}$ and the edges $(w_{j_{r+1},j}, u_{j_{r},j}),(u_{j_r,j}, u_{j_{r+1},j})$, for $r  \in[\gamma_j-2]$ and $(w_{j_1,j}, u_{j_1,j}),(w_{j_{\gamma_j},j}, u_{j_{\gamma_j-1},j})$ and $(u_{j_{\gamma_j-1},j}, s_j).$
 The \textit{reduction} gadget is presented in Fig.\ref{GraphDeltaEll} (b). \\
 \\
{\bf Path gadgets.} A path $P_j$  of $p = n +2nm$ nodes  departs from  each $s_j\in S$. 
See Fig.\ref{GraphDeltaEll} (c).
Notice that, by construction the degree of nodes is upper bounded by $3$.
 We set now the thresholds of the nodes in $G$ as: $t(v)=0$ for each node $v\in A$, $t(v)=2$ for each node $v\in U$ and $t(v)=1$ for all the remaining nodes.

\begin{lemma} \label{HS-IIB}
\small{{$\langle\{S_1,\ldots, S_m\}, A, h\rangle$}} is a {\sc yes} instance of HS iff
$\langle G, p,h\rangle$ is a {\sc yes} instance of \IIB.
\end{lemma}

\begin{proof}
Suppose that there exists $H \subseteq A$ such that $|H| \leq h$ and  $H \cap S_j \neq \emptyset$  for each $j\in [m]$. 
Consider in $G$ the set of nodes $Y$ corresponding to the elements of $H$.
Since $H \cap S_j \neq \emptyset,$  for each $j\in [m]$, we have that   each node $s_j\in S$  is connected, through a reduction gadget, with a node  in $w_{i,j}$ such that $a_i \in S_j\cap Y$.   Consequently, if all the nodes in $Y$ are immunized, then at least one node in the reduction gadget associated to $s_j$ cannot reach the threshold and consequently $s_j$ will not be influenced. Hence, no node in $S$  as well as  in the associated path gadgets can get influenced. 
We  have 
$|Y|\leq  h$ and $|\A_{G,Y}| <  p$, where the last inequality follows noticing that   $p=n+2nm$ is greater than the   number of  nodes that remain in $G$ once   we eliminate  the nodes in $S$ and in the  path gadgets.

Assume now there exists  a solution $Y$ to \IIB\  such that  $|Y|\leq h$ and $|\A_{G,Y}|\leq p$.
Without loss of generality, we can assume that $Y	\subseteq A$. Indeed, if $Y$ contains either of the nodes $w_{i,i_j}, u_{i,i_j}, s_{i_j}$ or a node in the path $P_{i_j}$, for some $i\in [n]$, we could replace such a node by  $a_i\in A$  without increasing neither the size of $Y$ nor  $\A_{G,Y}.$
Hence, we have that $Y$ consists of  at most  $h$ nodes in $A$.
We argue that the set $H\subseteq A$ of the elements corresponding to the nodes in $Y$ satisfies $H \cap S_j \neq \emptyset,$  for each $j\in [m]$.
Indeed, assume by contradiction that there is a set $S_j$ such that $H \cap S_j = \emptyset$. This implies that in   $G$  the node  $s_j$ will be influenced. Indeed, $s_j$  is connected  through    gadgets, to all the nodes in $S_j$. Moreover each node in $S_j$ belongs to $A-Y$ and has threshold $0$. It follows that  $s_j$ and, as a consequence, all the $p$ nodes on  the associated path get influenced and we obtain the desired contradiction because this violate the bound on the size of $\A_{G,Y}$.
\end{proof}

\subsection{Graphs of bounded treewidth} \label{secLBTW}
\begin{definition}\label{treedec}
A tree decomposition of a graph $G=(V,E)$ is a pair $(T,\{W_u\}_{u \in V(T)})$, where $T$ is a tree in which  each node $u$ is assigned a node subset $W_u \subseteq V$ such that:
\\
1.  $ \bigcup_{u \in V(T)} W_u=V$.
\\
2. For each edge $e=(v,w)\in E,$ there exists  $u$ in $T$ such that $W_u$ contains both $v$ and $w$.
\\
3. For each $v \in V$, the set $T_v=\{ u \in V(T) : v \in W_u\},$ induces a connected subtree of $T$.
\end{definition}
\noindent The width of a tree decomposition  $(T,\{W_u\}_{u \in V(T)})$ of a graph $G$, is $\max_{u \in V(T)}|W_u|-1$.
The treewidth of  $G$, denoted by $\tw(G)$, is the minimum  width of a tree decomposition of $G$.
\begin{theorem} \label{th:hard_tw}
\IIB\ is $W[1]$-hard with respect to the treewidth of the input graph. 
\end{theorem}
 In order to prove Theorem \ref{th:hard_tw}, we present a reduction from {\textsc{Multi-Colored clique (MQ)}}:
{\em Given   a graph $G=(V,E)$ and  a  proper vertex-coloring ${\bf c} : V \to [ q]$ for $G$, 
does $G$ contain  a  clique  of size $q$?}
\\
Given an instance $\langle G,q \rangle$ of  {\sc MQ}, we construct an instance $\langle G'=(V',E'),k,\ell\rangle$ of \IIB.
We denote by $n'=|V'|$ the number of nodes in $G'$. For a color $c \in [q]$, we denote by $V_c$ the class of nodes in $G$ of color $c$ and for a pair of distinct   $c,d \in [q],$ we let $\Ecd$ be the subset of edges in $G$ between a node in   $V_c$ and one in $V_d$.

\begin{figure}[tb!]
	\centering
		\includegraphics[width=0.9\linewidth]{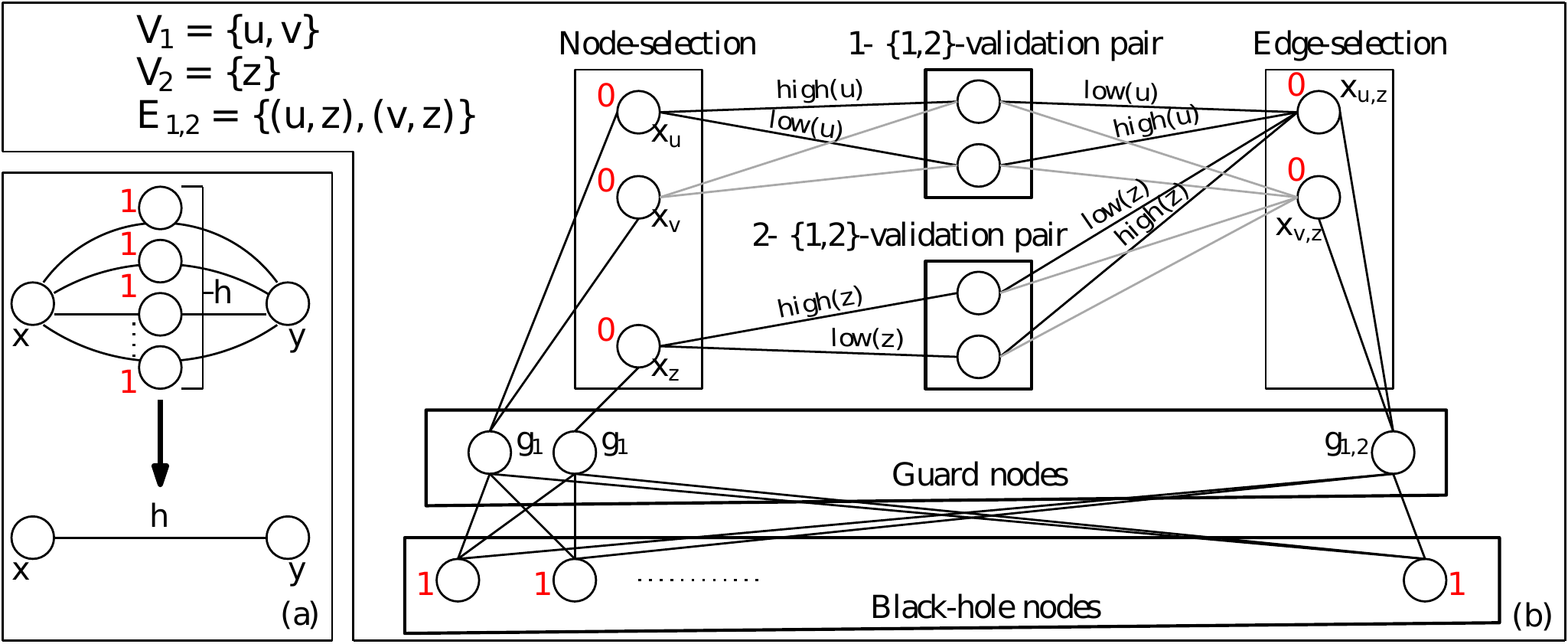}
	\caption{(a) Parallel-paths	gadget. (b) Representation of the graph $G'$ for a trivial instance of the  {\sc MQ} problem $\langle G=(V_1 \cup V_2, E_{1,2}), 2\rangle$. \label{fig:gprime}
	}
\end{figure}

Our goal is to guarantee that any solution of  \IIB\  in $G'$ encodes a clique in $G$ and vice-versa. Following some ideas in \cite{Ben-Zwi}, we construct $G'$ using the following gadgets:
\\
\\
\textbf{Parallel-paths gadget:} 
A parallel-paths gadget of size $h$, between nodes $x$ and $y$,  consists of   $h$ disjoint paths each made up by   a connection node which is adjacent to both  $x$ and $y$.   In order to avoid cluttering, we draw such a gadget as an  edge with label $h$ (cf. Fig. \ref{fig:gprime} (a)).
\\
\\
\noindent \textbf{Selection gadgets:} The selection gadgets encode the selection of   nodes (node-selection gadgets) and  edges (edge-selection gadgets):
\begin{itemize}
	\item[] {\bf Node-selection gadget:} For each $c \in [q]$, we construct a $c$-node-selection gadget which consists of a node $x_v$ for each $v \in V_c$; these nodes are referred as node-selection nodes. We then add a guard node $g_c$  that is connected to all the other nodes in the gadget; thus the gadget  is a star  centered at $g_c$. 
\item[] {\bf Edge-selection gadget:} For each   $c,d \in [q]$  with $c\neq d$, we construct a $\{c,d\}$-edge-selection gadget which consists of a node $x_{u,v}$ for every edge $(u,v) \in  \Ecd$; these nodes are referred as edge-selection nodes. We then add a guard node $g_{cd}$ that is connected to all the other nodes in the gadget; thus the gadget  is a star  centered at $g_{cd}$. 
	\end{itemize}
Overall there are  $n$ node-selection nodes with $q$ guard nodes and $m$ edge-selection nodes with ${q \choose 2}$ guard nodes (cf. Fig. \ref{fig:gprime} (b)). 
\\
\\
\textbf{Validation gadgets:} We assign to every node $v \in V(G)$ two unique identifier numbers, $low(v)$ and $high(v)$, with $low(v) \in [n]$ and $high(v)=2n-low(v)$.
	For every pair of distinct   $c,d \in [q],$ we construct two validation gadgets. One between the $c$-node-selection gadget and the $\{c,d\}$-edge-selection gadget  and one between the $d$-node-selection gadget and the $\{c,d\}$-edge-selection gadget.  
We describe the validation gadget between the $c$-node-selection and $\{c, d\}$-edge-selection gadgets.  It consists of two nodes. 
The first one 
is connected to each node $x_v$, for $v \in V_{c}$, by parallel-paths gadgets of size $high(v)$, and to each edge-selection node $x_{u,v},$ for $(u, v) \in \Ecd$ and $v \in V_{c}$,   by parallel-paths gadgets of size $low(v)$. The other node is connected to each node $x_v$, for $v \in V_{c}$,   by parallel-paths gadgets of size $low(v)$, and to each edge-selection node $x_{u,v},$ for $(u, v) \in \Ecd$ and $v \in V_{c}$, by parallel-paths gadgets of size $high(v)$.
Overall, there are $q (q-1)$ validation gadgets, each  composed by two  
nodes. 
\\
\\
\textbf{Black-hole gadget:} 
We add a set $B$ of $|B|=(n-q)(2nq-2n+1)+\left(m-{q \choose 2}\right)(4n+1)$ independent nodes and   a complete bipartite graph between  nodes in  $B$ and the  guard nodes.

To complete the construction, we specify the thresholds of the nodes in $G'$
$$t(x)=\begin{cases}
{0} &{\mbox{if $x$ is a selection node}}\\
{1} &\mbox{if $x$ is a connection node or $x\in B$}\\
{d_{G'}(x)-2n+1} &\mbox{if $x$ is a validation node}\\
{|V_c|} &\mbox{if $x=g_c$ is a guard node for some  $c \in [q]$ }\\
{|E_{cd}|} &\mbox{if $x=g_{cd}$ is a guard node for some  $c,d \in [q]$ }\\
\end{cases}$$
The complete construction of $G'$ for an instance of the {\sc MQ} problem   appears in Fig. \ref{fig:gprime} (b).

\begin{lemma} \label{MQ-IIB}
$\langle G, q\rangle$ is a {\sc yes} instance of MQ if and only if
$\langle G',k,\l\rangle$, where $k=(n-q)(2nq-2n+1)+\left(m-{q \choose 2}\right)(4n+1)$ and $\l= q+ {q \choose 2}$ 
is a {\sc yes} instance of \IIB. 
\end{lemma}

\begin{proof}
We first notice that  a  node $v$ can belong to the desired clique only if $\{v\}\cup \Gamma_{G}(v)$ contains at least one node from each color class.  Hence, we can remove from $G$ all   the nodes that do not satisfy such a property, since they are irrelevant to the problem.

Suppose that $K=(V(K),E(K))$ is a multi-colored clique in $G$ of size $q$.
Let $C$ denote the set of connection nodes and $X_K= \{x_v : v \notin V(K)\} \cup \{x_{u,v} : (u,v) \notin E(K)\}$. We set
$$X= X_K \cup \{c \in C : \Gamma_{G'}(c)\cap X_K\neq \emptyset\}.$$
We show that 
$$Y = \{x_v : v \in V(K)\} \cup \{x_{u,v} : (u, v) \in E(K)\}$$
 is the immunizing set  of $X$, i.e.,  $Y=Y(X)$. Notice that $|Y|= q + {q \choose 2}.$

We first observe that $\A_{G'[X]}=X.$ Indeed, nodes in $\{x_v : v \notin V(K)\} \cup \{x_{u,v} : (u,v) \notin E(K)\}$ have threshold $0$ and their neighbors in $C$ have threshold $1$.
Now we can easily evaluate the size of  $X$. Indeed $X$ is composed by:
\begin{itemize}
	\item $n-q$ nodes in the set of node-selection nodes and  their  $(n-q)2n(q-1)$  neighbors in $C$. Indeed,  each node-selection node is  connected with $q-1$ validation pair and, for each node $x_u$, we have  $low(u)+high(u)=2n$. 
	\item $m-{q \choose 2}$ nodes in the set of edge-selection nodes and their $(m-{q \choose 2})4n$ neighbors in $C$.  Indeed, each edge-selection node is connected with two validation pair and for each node $x_{u,v}$ we have that $low(u)+high(u)=low(v)+high(v)=2n.$
\end{itemize}
Overall the set $X$ has size
\begin{equation}\label{eqX}  
k=(n-q)(2nq-2n+1)+\left(m-{q \choose 2}\right)(4n+1).\end{equation}
 It remains  to show that $Y=Y(X)$. 
First of all, we observe that $Y\subseteq Y(X)$ because all the nodes in $Y$ belongs to $V'-X$ and  have threshold $0$, hence, by  (\ref{Y(X)}), each node in $Y$ belongs to $Y(X)$. We show now that  for any   $v \in V'- X$ it holds $|\Gamma_{G'}(v)\cap X|<t(v)$. 
\begin{itemize}
\item Each guard node $g$ has a neighbor in $Y$ and its threshold  is equal  to the number of its neighbors belonging to its selection gadget. Hence,  $|\Gamma_{G'}(g)\cap \A_{G'[X]}|<t(g)$. 
\item For each  $b\in B$, it holds    $|\Gamma_{G'}(b)\cap X|=0<t(b)=1.$ 
\item 
Consider now the validation nodes. Knowing that $K$ is a multi-colored clique, we have that for each validation pair there is exactly one node $u$ and one edge $(u,v)$ such that  $x_u, x_{u,v}\in Y$. Hence, both  nodes  have exactly $low(\cdot)+high(\cdot)=2n$ neighbors which do not belong to $X$. Since the threshold of each validation node $x$ is $t(x)=d_{G'}(x)-2n+1,$ then $|\Gamma_{G'}(x)\cap X|=d_{G'}(x)-2n<t(x).$ 
\item
Finally, for each connection node $c\notin X,$  we have $|\Gamma_{G'}(c)\cap X|=0<t(c)=1.$
\end{itemize}

Assume now there exists a solution $Y$ to \IIB \  such that $|Y| \leq \l=q+{q \choose 2}$ and
\begin{equation}\label{eqAk2}
|\A_{G',Y}| \leq k=(n-q)(2nq-2n+1)+\left(m-{q \choose 2}\right)(4n+1).
\end{equation}

Noticing that $k<|B|+1$ and all the nodes in $B$ get influences as soon as a guard node is, we have that the immunization of $Y$  saves all the guard nodes. Noticing that the number of guard nodes is exactly $q + {q \choose 2}$ and each guard node is connected to a separate set of selection nodes, we have that $|Y|=q + {q \choose 2}$ and each node in $Y$ can  save one guard node.
Recalling that the thresholds of guard nodes is equal to the number of neighbors belonging to the corresponding selection gadget, we have that in order to save a guard node there are two options: Put the guard node in $Y$ or put in $Y$ one of its neighbors,  belonging to the corresponding selection gadget.
Without loss of generality, we can assume that $Y$ does not include any guard node. Indeed, if $Y$ contains a guard node we could replace  such a node by one of its selection node neighbors without increasing neither the size of $Y$ nor of $\A_{G',Y}$.

We can then assume that $Y$ is composed by exactly $q$ node-selection nodes and $q\choose 2$ edge-selection nodes.
Let $V_Y\subseteq V$ be a set of $q$ nodes in $G$, defined by $V_Y = \{v \in V : x_v \in Y\}$.
We argue that $G[V_Y]$ is a clique.
By contradiction suppose that $G[V_Y]$ is not a clique. There are two nodes $u, v \in V_Y$ such that $(u, v) \notin E$. Let $c,d$ respectively the colors of $v$ and $u$. Let $x_{w,z}$ the node in $G'$ which save the guard $g_{cd}$ associated to the pair $c,d$. Since $(u, v) \notin E$ we have that $w \neq u$ or $z \neq v$ or both. Without loss of generality, we can assume that $w \neq u.$
Consider now the validation pair  between the $c$-node- and $\{c, d\}$-edge-selection gadgets. 
Recalling that $Y$ contains exactly one node for each selection gadget, we have that both the nodes in the validation pair have all the neighbors influenced,  except for the connections of the nodes $x_u$ and  $x_{w,z}$. Since $w \neq u$, we have that one of the vertices in the validation pair will get influenced. This is because for any $w \neq u$ either $high(w)+low(u) < 2n$ or $low(w)+high(u) < 2n$. That is, there is a validation node $x$ having less than $2n$ not influenced neighbors, while all the remaining neighbors get influenced. Recalling that the threshold of $x$ is $d_{G'}(x)-2n+1,$ we have that $x$ get influenced.

Hence, $|\A_{G',Y}|=k+1$. Indeed $k$ are due to non immunized selection nodes and their connection neighbors (see (\ref{eqX})) plus at least one validation node. This contradicts (\ref{eqAk2}).
\end{proof}

\begin{lemma}  \label{tw-G'}
$G'$ has treewidth $O(q^2)$.
\end{lemma}

\begin{proof}
We show now that $G'$ admits a tree decomposition of width $O(q^2)$.
The complete bipartite network defined by the guard nodes and the nodes in $B$ has treewidth  $q + {q \choose 2}$. Let $A$ be the set of the guard nodes of size $q + {q \choose 2}$  and  $b_1,b_2,\ldots,b_{\hat{n}}$ the  nodes in $B$. The decomposition tree has $A$ as root and $A\cup{b_i}$ as children. Then  we can  add to this network the $q + {q \choose 2}$ trees, rooted  on the guard nodes and containing both selections and connection nodes, without increasing the treewidth. Finally we can add all $O(q^2)$ validation nodes, getting a tree decomposition of width $O(q^2)$ for $G'.$
\end{proof}

\subsection{Graphs of bounded neighborhood diversity} \label{secLBND}
Given a graph $G=(V,E)$,  two nodes $u,v\in V$  are said to have the same  
{\em type} if $\Gamma_G(v) \setminus \{u\} = \Gamma_G(u) \setminus \{v\}$.
The {\em neighborhood diversity} of a graph $G$, introduced by Lampis in \cite{L} and denoted by {\em \nd}$(G)$, is the minimum number $\nd$ of sets in a partition  $V_1,V_2, \ldots, V_\nd$, of the node set $V$, 
such that all the nodes in  $V_i$ have the same type, 
for  $i\in [\nd]$.
The family 
$\{V_1,V_2, \ldots, V_\nd\}$ is called  the {\em type partition} of $G$.
\\
%
Notice that 
 each $V_i$ induces either a {\em clique} or an {\em independent set} in $G$.
Moreover, for each  $V_i,V_j$ in the type partition, we get that  either each node in 
$V_i$ is a neighbor of each node in $V_j$ or no node in $V_i$ 
has a neighbor  in $V_j$.
Hence, between each pair $V_i,V_j$, there is either a complete bipartite graph  
or  no edges at all.

\begin{theorem}\label{w-nd}
 \IIB\  is W[1]-hard with respect to the  neighborhood diversity of the input graph. 
\end{theorem}
In order to prove Theorem \ref{w-nd}, we use a reduction from {\sc Multi-Colored clique (MQ)}, defined in Section \ref{secLBTW}.
As before, we refer to  $\Vc$ 
as a color class of
$G$ and to  $\Ecd$ as the set of edges between nodes in the color classes 
$\Vc$ and $\Vd$. Here we will use the fact that  \MQ\  remains W[1]-hard 
even if  each color class  has the same size and for each {distinct colors} $c,d \in[\q]$, the set $\Ecd$  has the same size \cite{CFKLMPPS15}.
We then denote by $\r+1$ the size of each color class  
$V_c$
and by $\s + 1$ the size of each set $\Ecd$, in particular   we use the following notation
\begin{equation}\label{VE}
V_c =\{\vco,\vcu, \ldots, \vcr\},\  \qquad\   \Ecd=\{\ecdo, \ldots, \ecdm\}
\qquad  \mbox{ $c,d \in [q]$, $c\neq d$}
\end{equation}
and refer to $\vci$ and $ \ecdj$ as the $i$-th node in $V_c$ and the $j$-th edge in $\Ecd$, respectively.

Let $\langle G, \q \rangle$ be an instance  of \MQ. 
 We describe a reduction from  $\langle G, \q \rangle$  to an instance $\langle G', \k,  \l\rangle$ of \IIB\  such that $\nd(G')$ is $O(\q^2)$.
The reduction runs in time $poly(|G|)$.

\remove{Recall that in the \MQ\  problem we need to select exactly 
one node from each color class $\Vc$ and exactly one edge from each set 
$\Ecd$. Moreover, we have to make certain
that if $(u, v) \in \Ecd$ is a selected edge, then $u \in \Vc$ and $v \in \Vd$ are selected nodes.}

In order to present the reduction we introduce some gadgets that are used in the construction of  $G'$. They are inspired by those used in \cite{DKT16}.
The rationale behind the construction is the following. First, we create two sets of gadgets (Selection and Multiple gadgets), which encode in  $G'$ the selection of nodes and edges as part of  a potential multicolored clique in $G$. Then we create another set of gadgets (Incidence gadgets) that is used to check whether the selected sets of nodes and edges actually  represent a multicolored clique in $G$. Our goal is to guarantee that any solution of \IIB\ in $G'$ encodes a clique in $G$ and vice-versa.

\begin{figure}[tb]
	\centering
	\includegraphics[width=0.85\textwidth]{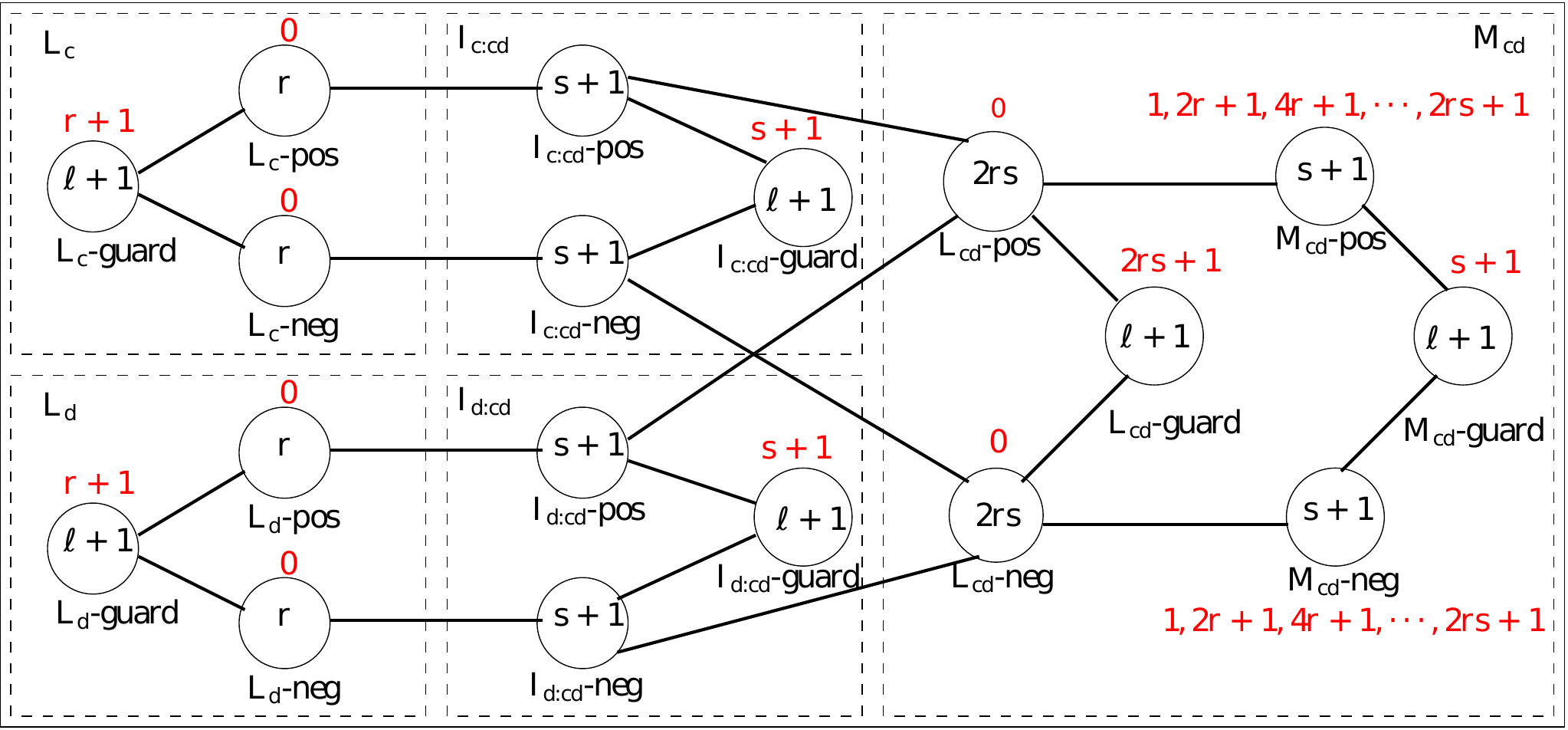}
	\caption{\small An overview of the reduction. Each circle represents a \bug. The number inside a \bug\ is the number of nodes of
the \bug. The threshold of nodes in a \bug\ is displayed in red.
 }
	\label{reduction-nd}
\end{figure}

In the following we  call {\em \bug} 
an independent set of nodes of a graph sharing all neighbors.
So, a connection between two \bug s  points out a complete bipartite graph among the nodes in the  \bug s.
Fig. \ref{reduction-nd} shows the gadgets we are going to introduce and how they are connected.

\medskip
\noindent
{\bf Selection Gadget.}
For each  $c \in [\q]$, 
{the selection gadget $\Lc$ consists of three \bug s: $\Lcn$ and $\Lcp$ of $\r $ nodes each, and $\Lcg$
of $\l+1$ nodes (the value $\l$, representing an upper bound on the number of nodes to be immunized, will be determined later). The \bug\ $\Lcg$  is connected to both $\Lcn$ and $\Lcp$.
 We set the threshold  of each node $g $ in $ \Lcg$ to $t(g)=\r+1$ and  the threshold
of each node $v $ in $\Lcn\cup \Lcp$   to $t(v)=0$. }
The selection gadget $\Lc$ is connected to the rest of the graph $G'$ using only nodes from $\Lcn\cup \Lcp$.
\\
\\
{\bf Multiple Gadget.}
For each $c, d \in [q]$ with $c\neq d$, 
{we create a multiple gadget $\Mcd$ 
consisting of six \bug s: $\Lcdp$ and $\Lcdn$ of $2\r \s$ nodes each, $\Lcdg$ of $\l+1$ nodes, $\Mcdp$ and $\Mcdn$ of $\s+1$ nodes each, and 
$\Mcdg$ of $\l+1$ nodes. 
$\Mcdg$  is connected to the  \bug s $\Mcdp$ and $\Mcdn$.
$\Mcdp$ is connected to $\Lcdp$, and $\Mcdn$ is connected to $\Lcdn$.
Finally, the \bug\  $\Lcdg$ is connected to both $\Lcdp$ and $\Lcdn$.
The rest of graph $G'$ is connected only to the \bug s
$\Lcdp$ and $\Lcdn$.
We set the threshold of each $g \in \Mcdg$ to 
$t(g)=\s+1$.
For each node $v\in\Lcdp \cup \Lcdn$, we set 
the threshold $t(v)=0$.}
\remove{This means that node $v$ is influenced at the first round of the diffusion process in $G'$; hence, it has to be a node in $X \cup Y$.}
Let $\Mcdp=\{x_0, \ldots, x_\s\}$ and $\Mcdn=\{y_0, \ldots, y_\s\}$;
we set thresholds $t(x_i) = t(y_i) = 2\r i+1$. 
Finally, for each  $g \in \Lcdg$, we set 
the threshold $t(g)=2rs+1$.

\medskip
\noindent
{\bf Incidence Gadget.} For each pair of distinct  $c,d \in [q]$, we construct two incidence gadgets: $\Iccd$ (connected with the gadgets $\Lc$ and $\Mcd$) and $\Idcd$ (connected with the gadgets $\Ld$ and $\Mcd$). In the following we present the gadget $\Iccd$ which has the same structure of the gadget $\Idcd$.
The incidence gadget $\Iccd$  has three \bug s $\Iccdp$ and $\Iccdn$
of $\s + 1$ nodes each, and $\Iccdg$ of $\l+1$ nodes.
We connect $\Iccdg$ to $\Iccdp$ and $\Iccdn$. 
Furthermore, we connect $\Iccdp$ to $\Lcp$ and 
$\Lcdp$. Similarly, we connect  $\Iccdn$ to  
$\Lcn$ and $\Lcdn$.
We set the threshold of each $g \in \Iccdg$ to $t(g)=\s + 1$.
Recalling that there are $\s + 1$ edges in the set $\Ecd$, and 
that there are $\s+1$ nodes in $\Iccdp$ and $\Iccdn$,
we create  one-to-one correspondences  between $\Ecd$ and $\Iccdp$
and between $\Ecd$ and $\Iccdn$.
Namely, for each $j=0,\ldots  s$, we associate the  $j$-th edge $\ecdj$  in $\Ecd$ (cfr. (\ref{VE})) to a   node  $u_j \in \Iccdp$ and to a node $w_j \in \Iccdn$ (with $u_j\neq u_{j'}$ and $w_j\neq w_{j'}$, for $j\neq j'$). Moreover, if the endpoint of $\ecdj$ of color $c$ is   the $i$th node  $\vci$ of $\Vc$ (cfr. (\ref{VE})) then 
we set 
 \centerline{$t(u_j)= i +1 + 2\r j,\qquad t(w_j)= \r -i +1+ 2\r (\s -j).$} \\
It is worth  observing that the  nodes in $\Iccd$-pos (respectively, $\Iccd$-neg) have  different thresholds. Indeed,  the numbers $i +1+ 2\r j$ (respectively, $\r -i +1+ 2\r (\s -j)$) are all  different, for $0\leq i \leq r$ and $0\leq j \leq s$.

\medskip
\noindent
{\bf Black-hole Gadget.}
Finally we add a gadget, which will force the immunizing set $Y$ to contain a specific number of nodes for selection ($r$ nodes) and multiple gadgets ($2rs$ nodes). We add a bag $B$ of $|B|=\q \r +   \binom{\q}{2} (2\r + 3) \s$  nodes and  connect it to the guard bags  in all the
selection, multiple and incidence gadgets.
For each $v \in B$, we set  $t(v)=1$.

\begin{lemma} \label{clique<->immuno}
$\langle G, q\rangle$ is a {\sc yes} instance of \MQ\ iff
$\langle G', k,\l\rangle$
is a {\sc yes} instance of \IIB,
where $k= \q \r +   \binom{\q}{2} (2\r + 3) \s$ and 
$\l = \q \r +  \binom{\q}{2} 2\r\s.$ 
\end{lemma}

The proof of Lemma \ref{clique<->immuno} will follow by
 Claims \ref{clique-immuno},   \ref{immuno-clique}  proved below.
 
\begin{claim} \label{clique-immuno}
If $\langle G, q\rangle$ is a {\sc yes} instance of \MQ\ then
$\langle G', k,\l\rangle$
is a {\sc yes} instance of \IIB.
\end{claim}
\proof
Let $K=(V(K),E(K))$ be  a multicolored clique of $G$.
We will show how to select nodes to be  added to the immunizing set $Y$ according to  the nodes in  $K$.
First of all notice that,  all the nodes in the \bug s
$\Lcp$, $\Lcn$, $\Lcdp$, and $\Lcdn$ belong to  $Y \cup \A_{G',Y}$,
as they all have threshold zero.

For each   $c \in [q],$  if the unique node of color $c$ in $K$ is   $\vci$, the $i$-th node in $V_c$, then 
we add $i$ nodes of $\Lcn$  and $\r-i$ nodes of  $\Lcp$ to $Y$.
For each pair of distinct   $c,d \in [q],$ 
if the unique edge with endpoints of colors $c$ and $d$ in $K$  is $\ecdj$, then 
we add $2\r j$ nodes of $\Lcdn$ and $2\r (\s -j)$ nodes of $\Lcdp$ to $Y$. 
Overall, $|Y|= \ell=\q \r +  \binom{\q}{2} 2\r \s$.
We now prove that $|\Active_{G',Y}|=k=\q \r +   \binom{\q}{2}  (2\r + 3) \s$.

Consider the diffusion process in $V(G')-Y$.
At the first round, all non immunized nodes with threshold zero are influenced; hence $\Active_{G',Y}[1]$ contains: $i$ nodes of $\Lcp$, for all   $c\in [\q]$ and
$\r -i$ nodes of $\Lcn$,   $2\r j$ nodes of $\Lcdp$, $2\r (\s -j)$ nodes of $\Lcdn$, for all   $c,d\in [\q]$ with $c\neq d$.

We claim that, at the second round, the additional influenced nodes (in the neighborhood of  $\Active_{G',Y}[1]$) are exactly:  $\s$ nodes in $\Mcdp \cup \Mcdn$,  $\s$ nodes in $\Iccdp \cup \Iccdn$,  and $\s$ nodes in $\Idcdp \cup \Idcdn$, for each pair of distinct  $c,d \in [\q]$. 
Indeed, let  $\Mcdp = \{x_0, \ldots, x_\s \}$
and $\Mcdn=\{y_0, \ldots, y_\s\}$. 
Since at the end of the first round the nodes in $\Mcdp$ have $2\r j$ influenced neighbors in $\Lcdp$ and the nodes in $\Mcdn$ have $2\r (\s -j)$ influenced neighbors in $\Lcdn$, recalling that $t(x_j)=t(y_j)=2\r j+1$, we have that nodes $x_0, \ldots, x_{j-1}$ in $\Mcdp$ and nodes $y_0, \ldots, y_{\s -j-1}$  in $\Mcdn$  get influenced. Overall $\s$ nodes in $\Mcdp \cup \Mcdn$ are influenced at the second round.\\
Consider now the incidence gadgets.
Since there are $2\r j+i$ influenced nodes in  
$\Lcp \cup\Lcdp$ that are in neighborhood of the nodes in $\Iccdp$, recalling that the thresholds of nodes in $\Iccdp$ are: 
\begin{eqnarray*}
t(u_j)&=& 2\r j+i+1>2\r j+i \mbox{ and}\\
t(u_h)&=& 2\r h+h'+1  \mbox{ for each $0 \leq h \leq s, h\neq j$, and   $0 \leq h'\leq \r$},
\end{eqnarray*}
we have 
\begin{eqnarray*}
t(u_h) &\leq& 2\r h+\r +1 \leq 2\r (j-1)+r+1 =2\r j -\r +1\leq 2\r j +i  \quad
\mbox{if $h < j$}\\
t(u_h)&\geq& 2\r h +1 \geq 2\r (j+1)+1 >2\r  j+2\r +1 >2\r j+i  \qquad\qquad 
\mbox{if $h > j$.}
\end{eqnarray*}
Hence, nodes $u_0, \ldots, u_{j-1}$ in $\Iccdp$ are influenced at the second round.\\
We now make a similar analysis for the nodes in $\Iccdn$.
Since there are $\r -i+2\r (\s-j)$ influenced nodes in  
$\Lcn \cup \Lcdn$  that are in neighborhood of the nodes in $\Iccdn$, recalling that the threshold of nodes in $\Iccdp$ are: 
\begin{eqnarray*}
t(w_j)&=& 2\r (\s-j)+\r -i+1 \ > \ 2\r (\s-j)+\r -i \mbox{ and}\\
 t(w_h) &=&  2\r (\s-h)+\r -h'+1 \mbox{ for some $0 \leq h'\leq \r$},
\end{eqnarray*}
we have
\begin{eqnarray*}
&&t(w_h) 
 \geq  2\r (\s-h)+1 \geq 2\r (\s-j)+2\r +1 > 2\r (\s-j)+\r -i \qquad 
\mbox{for $h < j$}\\
&&t(w_h)   \leq 2\r (\s-h)+n+1 \leq 2\r (\s-j)-\r +1 \leq 2\r (\s-j)+\r -i 
\quad \mbox{for $h > j$}.
\end{eqnarray*}
Hence, nodes $w_{j+1}, \ldots, w_\s$ in $\Iccdn$ are influenced at the second round. 
Overall,  we have that  $s$ nodes in  $\Iccdp \cup\Iccdn$ are influenced at the second round.\\
Using exactly the same argument we can show that $s$ nodes in  $\Idcdp \cup\Idcdn$ are influenced at the second round. 

Finally, the nodes in  $\Lcg$ (resp. $\Lcdg$) have $\r$ (resp. $2\r \s$) influenced neighbors at the end of the first round and since all of them have threshold $r+1$ (resp. $2\r \s+1$), we have that none of them gets influenced at the second round.

We notice now that only the nodes in $\Mcdg$ and $\Iccdg$  have 
neighbors in $\Active_{G',Y}[2]$. However,   they cannot be influenced  (indeed, each of them has threshold $\s+1$ but it has  only $s$ influenced neighbors in 
$\Active_{G',Y}[2]$ -- in $\Mcdp\cup  \Mcdn$ or  in $\Iccdp \cup\Iccdn$). We have that
$\Active_{G',Y}[3]=\Active_{G',Y}[2]$ and  the diffusion process stops.

Summarizing, $\Active_{G',Y}$ contains: $\r$ influenced nodes for each 
of the $\q$ nodes in the clique $K$ (those  that are 
influenced in the selection gadgets $\Lc$ for $c\in[\q]$),   
$2\r \s +\s$ influenced nodes for each of the 
$\binom{\q}{2}$  edges in $K$ (those in the multiple gadgets $\Mcd$, for $c,d \in[\q]$) and $2s$ influenced nodes, for each of the 
$\binom{\q}{2}$  edges in $K$ (those in the incidence gadgets $\Iccd$ and $\Idcd$, for distinct $c,d \in[\q]$).
Hence, the set $\Active_{G',Y}$ contains $\k= \q \r +   \binom{\q}{2}  (2\r + 3) \s $ nodes.
\qed

\smallskip

Let $Y$ be an immunizing set such that $|Y| \leq \l=\q \r +  \binom{\q}{2} 2\r \s$ and $|\Active_{G',Y}|\leq k= \q \r +   \binom{\q}{2}  (2\r + 3) \s$.
In the following we derive some useful constraints on the nodes contained in $Y$ and $\Active_{G',Y}$.
\begin{prop}\label{guardie}
For distinct $c,d \in [\q]$, no node in $\Lcg$, $\Lcdg$, $\Iccdg$, $\Idcdg$, $\Mcdg$ can be in $\A_{G',Y}$.
\end{prop}
\proof
Since the threshold of each $v \in B$ is $t(v)=1$, it is sufficient that at least one guard node $g \in \Lcg \cup \Lcdg \cup \Iccdg \cup \Idcdg \cup \Mcdg$ is influenced to influence the whole $B$. However this cannot be since $|B|+1=k+1 >  |\Active_{G',Y}|$.
\qed
\begin{prop}\label{Lc-Lcd}
For distinct $c,d \in [\q]$, both $Y$  and $\A_{G',Y}$ contain \\
(1) \ exactly $\r$ nodes of $(\Lcp \cup \Lcn)$,\\
(2) \ exactly $2\r \s$ nodes of $(\Lcdp \cup \Lcdn$),\\
(3) \ a multiple of $2\r$ nodes of $\Lcdp$ and $\Lcdn$.
\end{prop}
\proof
First of all consider that all the nodes in $\Lcp$, $\Lcn$, 
$\Lcdp$ and $\Lcdn$ have threshold zero, and so
 all of them are in $Y \cup \Active_{G',Y}$.
We claim that at most $\r$ of the nodes of 
$(\Lcp \cup \Lcn)$ can be in $\A_{G',Y}$. Indeed, if $\A_{G',Y}$ contains at least $\r +1$ nodes in 
$(\Lcp \cup \Lcn)$ then each node  $g \in \Lcg$  (recall $t(g)=\r +1$)  either is influenced (i.e., $g \in \A_{G',Y}$) or is immunized (i.e., $g \in Y$).
By Proposition \ref{guardie}, no node in $\Lcg$ can be influenced.
On the other hand,  it cannot occur that all the nodes in $\Lcg$ are immunized, since $|\Lcg|=\l+1 > |Y|$.
\\
Using the same argument we can prove that at most $2\r \s$ of the nodes of $(\Lcdp \cup \Lcdn)$ can be in $\A_{G',Y}$. 
Assume on the contrary that $|\A_{G',Y}\cap (\Lcdp \cup \Lcdn)|\geq 2\r \s +1$. Having  each node in  $ \Lcdg$   threshold $2\r \s +1$, we have that either the node is influenced  or it must be  immunized. However,  by Proposition \ref{guardie} we know that  the  nodes in  $\Lcdg$ are not influenced; moreover they cannot all be immunized since   $|\Lcdg|=\l+1>|Y|$. 

This allows to say that 
$Y$ contains at least $\r$ nodes of 
$(\Lcp \cup \Lcn)$ and at least $2\r \s$ nodes of 
$(\Lcdp \cup \Lcdn)$.
However, if there exists a $c \in [\q]$ 
or a pair of distinct $c,d \in [\q]$ such that $Y$ contains strictly more than 
$\r$ nodes of $(\Lcp \cup\Lcn)$ or 
$2\r \s$ nodes of $(\Lcdp \cup \Lcdn)$, then
$|Y|>\q \r +  \binom{\q}{2} 2\r \s$  and this is not possible.
Hence, (1) and (2) follow.

To prove (3) we proceed by contradiction. Suppose that $\A_{G',Y}$ contains 
$2\r a+z$ nodes of $\Lcdp$, where $a < \s$ and $0 < z < 2\r$.
By (2) we have that  $\A_{G',Y}$ contains 
$2\r (\s-a)-z$ nodes of $\Lcdn$.
Write  $\Mcdp=\{x_0, \ldots, x_\s\}$ and $\Mcdn=\{y_0, \ldots, y_\s\}$.
Recalling that the nodes in $\Mcdp$  are neighbors of those in 
$\Lcdp$, the nodes in $\Mcdn$  are neighbors of 
those in $\Lcdn$  and  
$t(x_i) = t(y_i) = 2\r i+1$,
we have that nodes $x_0, \ldots, x_a$ of  $\Mcdp$
and nodes $y_0, \ldots, y_{\s-a-1}$ of  $\Mcdn$ get influenced.
Since these  $\s+1$ influenced nodes are neighbors of each node 
$g \in \Mcdg$,  whose threshold is $t(g)=\s+1$, we have that either $g$ is influenced or it is immunized.
By Proposition \ref{guardie}, no node in $\Mcdg$ can be influenced.
On the other hand,  it cannot occur that all the nodes in $\Mcdg$ are immunized, since $|\Mcdg|=\l+1 > |Y|$.
\qed

\begin{claim} \label{immuno-clique}
If $\langle G',k,\l\rangle$
is a {\sc yes} instance of \IIB\ then
 $\langle G, q\rangle$ is a {\sc yes} instance of MQ. 
%
\end{claim}
\proof
Being  $\langle G',k,\l\rangle$  a {\sc yes} instance of \IIB,  there exists an immunizing set $Y$ of size at most $\l=\q \r +  \binom{\q}{2} 2\r\s$ such that
$|\A_{G',Y}| \leq k=\q \r +   \binom{\q}{2} (2\r + 3) \s.$

We proceed by identifying  the  clique $K$ of $G$
according to the number of nodes that are in $\Lcn \cap Y$ for each $c\in [q]$
and in $\Lcdn \cap Y$, for each distinct $c,d \in [q]$ . Namely,
we select:

-- the node 
$\vci \in \Vc$, such that  $|\Lcn \cap Y|=i$, for some $0 \leq i \leq \r$,
and 

-- the edge 
$\ecdj \in \Ecd$ such that $|\Lcdn \cap Y|=2\r j$, for some $0 \leq j \leq  \s$.

\noindent
The above selection is correct since, by Proposition \ref{Lc-Lcd}, we know that $|Y\cap (\Lcp \cup \Lcn)|=\r$  and $|Y\cap (\Lcdp \cup \Lcdn)|= 2\r \s$ (in particular, $Y$ contains a multiple of $2\r$ nodes of both $\Lcdp$ and $\Lcdn$).

Let $V(K)$ be the set of the $q$ selected nodes and $E(K)$ be the set of the $\binom{\q}{2}$ selected edges.
We argue that $K=(V(K),E(K))$ is a clique.
By contradiction assume there are two distinct colors  $c,d \in  [q]$ such that  $\vci\in V(K)$ and $\ecdj\in E(K)$ but   $\vci$ is not an endpoint of $\ecdj$. 
Consider the incidence gadget $\Iccd$.
Let $\Iccdp = \{u_0, \ldots, u_\s\}$ and 
$\Iccdn = \{w_0, \ldots, w_\s\}$.
Assume that $\vch$ is the endpoint of color $c$ of $\ecdj$. Recall that nodes $u_j$ and $w_j$ represent the edge $\ecdj$ and that, by the construction of $G'$, it holds
$t(u_j)= 2\r j +h+1$ and $t(w_j)= 2\r (\s-j) +\r -h +1$.
Since the nodes of $\Iccdp$ have $2\r j+i$ influenced neighbors
(those in $\A_{G',Y} \cap (\Lcp \cup \Lcdp)$)
and the nodes of $\Iccdn$ have $2\r (\s-j)+\r -i$ influenced neighbors,
(those in $\A_{G',Y} \cap (\Lcn \cup \Lcdn)$)
by an analysis similar to that in the proof of 
Lemma \ref{clique-immuno},
we have that nodes $u_0, \ldots, u_{j-1}$ in $\Iccdp$  
and nodes $w_{j+1}, \ldots, w_\s$ in $\Iccdn$ all get
influenced.
It remains to analyze the nodes $u_j$ and $w_j$.
We will prove that  at least one of them gets influenced:
If $h < i$ then $t(u_j)= 2\r j +h+1 \leq 2\r j +i$ and
$t(w_j)= 2\r (\s-j) +\r -h +1 > 2\r (\s-j) +\r -i$ and 
 $u_j$ is influenced;
if $h > i$ then $t(u_j)= 2\r j +h+1 > 2\r j +i$ and
$t(w_j)= 2n(\s-j) +n-h +1 \leq 2\r (\s-j) +\r -i$ 
and $w_j$ is influenced.
This allows to say that if $\vch \in \ecdj$ then $\s+1$ nodes among those in 
$\Iccdp$ and $\Iccdn$ are influenced. 
As a consequence, 
each node 
$g \in \Iccdg$,  whose threshold is $t(g)=\s+1$,  must either be influenced or  immunized.
By Proposition \ref{guardie}, no node in $\Iccdg$ can be influenced.
On the other hand,  it cannot occur that all the nodes in $\Iccdg$ are immunized, since $|\Iccdg|=\l+1 > |Y|$.
\qed

\begin{lemma}
$G'$ has neighborhood diversity $O(q^2)$.
\end{lemma}
\begin{proof}
Since each bag in $G'$ is a type set in the type partition of $G'$ and,  since for each $c\in [q]$, there are three bags in $\Lc$ and, for each $c, d \in [q]$ with $c\neq d$ there are
six bags in $\Mcd$, and three bags in both $\Iccd$ and $\Idcd$,
we have that the neighborhood diversity of $G'$ is $3 q + 12 \binom{q}{2}$.
\end{proof}

\section{FPT Algorithms}\label{Algo}
In this section, we present FPT algorithm for several pairs of parameters.
\subsection{Parameters $k$ and  $\l$}\label{Algokl}
\begin{theorem}
\IIB\ can be solved in time $2^{k+\l}(k+\l)^{O(\log(k+\l))}\cdot n^{O(1)}$. \label{Th_k_l}
\end{theorem}

\begin{proof}
 The fixed parameter tractability of \IIB\ with respect to $k+\l$ can be proved by the  arguments  used in  Theorem 1 in \cite{Fomin}  for the problem {\sc cutting at most $k$ vertices with terminal}.
For sake of completeness, the complete proof is given in the following.   

Let $\langle G,k, \l \rangle$ be the input instance of \IIB. Consider a random labelling  of  the nodes of  $G$, where each node is independently assigned either 0 or 1 with equal probability. 
Let  now  $H=G[V_1]$ be the graph  induced by the set $V_1$ of  nodes having label 1. 
Consider the set $\Active_H$ of influenced nodes when we run the diffusion process on $H$. If
$|\Active_H|\leq k$ and $|Y(\Active_H)|\leq \l$ then (\ref{equiv}) holds for $X=\Active_H$ and  we can answer {\sc yes}.

We estimate now the number of  needed iterations of random labelling. Suppose   $G$  contains  a set $X$ satisfying (\ref{equiv}).
For such a set, it holds  $|X|=|\Active_{G[X]}|\leq k$ and  $|Y(X)|\leq \l$, then  a random labelling identifies a solution of \IIB\       if and only if    all the nodes in $X$ are labelled 1  and   all the nodes in $Y(X)$ are labelled 0, that is,  
  $$X\subseteq V_1  \mbox{ and } Y(X)\cap V_1=\emptyset.$$
Indeed, in such a case the above procedure identifies   $\Active_H=X$ as a solution.
This happens with   probability   $2^{-(|\Active_H|+|Y(\Active_H)|)}\geq 2^{-(k+\l)}$. Hence, the   algorithm requires time  $2^{k+\l} n^{O(1)}$.

A derandomization of the  above  process can be done using universal sets.
A $(n, i)$-universal set   is a collection of
binary vectors of length $n$ such that for each set of  $i$ indices,   each of  the $2^{i}$
possible combinations of values appears in some vector of the set. To run the algorithm, it suffices to try all labellings induced by a $(n, k + \l)$-universal set.
 Naor et al. [18] give  a construction of  $(n, i)$-universal sets of size $2^i i^{O(\log i)} \log n$ that can be
listed in linear time.
\end{proof}

\subsection{Parameters $k$ and  $\zeta$}\label{Algozeta}
\begin{theorem}
  \IIB\ can be solved in time   {$O(\zeta^{3k}n^5)$}, where   $\zeta=|\{v\in V\ |\ t(v)=0\}|$.
\end{theorem}
\proof 
Let $\langle G, k,  \l\rangle$ be the input instance of \IIB. 
Suppose  $v_1,\ldots v_\zeta$ are the nodes in $G$ having threshold 0 and let $\Delta$ denote the maximum degree of a node in $G$.
Consider the graph $G'=(V',E')$ obtained from $G$ by adding the internal nodes and the edges of  a $\Delta$-ry tree whose leaves are $v_1,\ldots v_\zeta$.
Assume  $\langle G, k,\l, \rangle$ is a {\sc yes} instance of \IIB. We notice that in $G$, the solution  set $X$ (cfr. (\ref{equiv})) can be 
disconnected but any of its connected components must include at least one node of threshold 0.  Hence,  in $G'$ the nodes in $X$  are now connected through a path in the $\Delta$-ry tree.   This implies that there exists $X'\subseteq V'$ such that: $X\subseteq X'$, $(X'-X)\subseteq V'-V$, 
and 
$G'[X']$ is connected. 
In particular, 
if $s$ is the root of  tree, we can assume  that $s\in X'$. In the worst case, all the paths within the $\Delta$-ry tree  go through the root $s$, hence  $|X'|\leq |X|\log_{\Delta} \zeta +1$.

Let $k'=k\log_{\Delta} \zeta+1$.
We use the following result [\citealp{subgraphs}, Lemma 2]: 
{\em 
There are at most $4^{k'}\Delta^{k'}$ connected subgraphs  that contain $s$ and have order at most $k'$. Furthermore, these subgraphs can be
enumerated in $O(4^{k'}\Delta^{k'}(|V'|+ |E'|))$ time.}
We can then apply the result in \cite{subgraphs} to enumerate all the connected subgraphs of $G'$ of size up to $k'$.
For each candidate set $X'$ (the node set of the current connected subgraph) one has to determine whether $X'\cap V$ is a  solution according to (\ref{equiv}),  which can be done in $O(n^2)$ time.
\hss\qed

\subsection{Parameters $k$ (or $\Delta$) and Treewidth}

In this section we present a dynamic programming algorithm which exploiting the tree decomposition of a graph $G$ enables to solve a minimization version of \IIB, namely the 

\begin{quote}
{\bf \textsc{{Influence Diffusion Minimization (IDM)}}}:
Given a graph $G=(V,E,t)$  and a  budget $\l$,
find a set $Y$ such that $|Y|\leq \ell$ and $|\Active_{G,Y}|$ is minimized.

\end{quote}

We  use the  rooted tree decomposition named nice tree decomposition.
\begin{definition}
\label{nice}
A 
tree decomposition $(T,\{W_u\}_{u \in V(T)})$ is nice if  conditions 1. and 2. hold:
\\
1. $W_r=\emptyset$ for $r$ the root of $T$ and $W_v=\emptyset$ for every leaf $v$ of $T$.
\\
2. Every non-leaf node of $T$ is of one of the following three types: 
\begin{description}
	\item[] Introduce: a node $u$ with exactly one child $u'$ such that $W_u=W_{u'}\cup \{v\}$ for a node $v\notin W_{u'}$.
	\item[] Forget: a node $u$ with exactly one child $u'$ such that  $W_{u'}=W_u\cup \{v\}$ for a node $v\notin W_{u}$.
	\item[] Join: a node $u$ with two children $u_1,u_2$ such that $W_{u}=W_{u_1}=W_{u_2}$
\end{description}
\end{definition}

\begin{lemma}\cite{Treewidth_book} \label{lemmanice}
If a graph $G$ admits a tree decomposition of width at most $\tw$, then it  admits a nice tree decomposition of width at most $\tw$. Moreover, given a tree decomposition $(T,\{W_u\}_{u \in V(T)})$ of $G$ of width at most $\tw$, one can  compute in time $O(\tw^2 \max\{|V(T)|,|V(G)|\})$ a nice tree decomposition of $G$ of width at most $\tw$ that has at most $O(\tw|V(G)|)$ nodes.
\end{lemma}

Consider a graph  $G=(V,E)$  with treewidth  $\tw$ and  nice tree decomposition $(T,\{W_u\}_{u \in V(T)})$.
Let  $T$ be rooted at  node $r$ and  denote by $T(u)$ the subtree of $T$ rooted at  $u$, for any node $u$ of $T$. Moreover, denote by  $W(u)$  the union of all the bags  in $T(u)$, i.e., $W(u)=\bigcup_{v \in T(u)} W_v$.  We will denote by  $s_u = |W_u|$ the size of $W_u$.

\def\C{{\cal C}}
\def\T{{\cal T}}

\def\CC{{\mathbb C}}
\def\TT{{\mathbb T}}

\def\xx{{0}}
\def\yy{{1}}
\def\zzz{{2}}
    
\newcommand{\XWG}[2]{S_{#1}(#2)}
\newcommand{\Tminus}[2]{#1(#2)}

We are going to recursively compute  the solution of  {\sc IDM}. The algorithm exploits a dynamic programming strategy and  traverses the input tree $T$ in a 
breadth-first fashion.
Moreover, in order to be able to recursively reconstruct the solution, we  calculate  optimal solutions under different
hypothesis based on the following considerations:\\
-- Fix a node $u $ in $ T,$ for each node $v \in W_u$ we have three cases:   $v $  gets influenced,  $v $ is immunized, or  $v$ is safe. 
We are going to consider all the $3^{s_u}$ combinations of such states. We  denote each  combination with a vector $\C $ of size $s_u$  indexed by the  elements of $W_u$, where the 
element indexed by $v\in W_u$
 denotes the state influenced ($\xx$), immunized ($\yy$), safe ($\zzz$) of node  $v$. 
{The configuration $\C=\emptyset$ denotes the vector of length 0 corresponding to an empty bag.} 
We  denote by $\CC_u$ the family of all the $3^{s_u}$ possible state vectors of the $s_u$ nodes in $W_u$. \\ 
--  Let $U$ be a subset of $V(G)$.
Let us first notice that by 3) of Definition \ref{treedec}, 
all the edges between nodes in $V-W(u)$ and $W(u)$ connect a node in $V-W(u)$ with a node in $W_u$ (the bag corresponding to the root of $T(u)$).
We are going to consider all the possible contribution to the diffusion process, of nodes in $V-W(u)$;  that is, for each $v \in W_u$, we consider all the possible residual thresholds among $t(v),t(v)-1,\ldots,\max\{0,t(v)-k\}$ (recall that at most $k$ nodes belong to $X$ and can therefore reduce the threshold of $v$). 
We notice that, for each node $v$, it is possible to bound the number of residual thresholds by  the value $\min\{t(v),k\}$. 
Moreover, since no node with $t(v)>d_G(v)$ can be influenced and can be then purged from $G$ in a preprocessing step, we can assume that in $G$ it holds $(\max_{v\in V} t(v))\leq \Delta$. Hence, we will have up to $\mu^{s_u}$ threshold combinations, where $\mu=\min\{\k,\Delta\}$.
We will denote each possible threshold combination with a vector $\T$, indexed by the  $s_u$ elements in $W_u$, where the 
element indexed by $v$ belongs to  $\{\max\{0,t(v)-k\},\ldots,t(v)\}$ and denotes the residual threshold of  $v \in W_u$. The configuration $\T=\emptyset$ denotes the vector of length 0 corresponding to an empty bag.
We  denote by $\TT_u$ the family of all the possible threshold combinations of nodes in $W_u$. 
 
The following definition introduces the values that will be computed by the algorithm in order to keep track of all the above cases:

\begin{definition}\label{eq_Xu} 
For each node $u\in T,$ each $j=0,\ldots, \l$, $\C\in \CC_u$  and  $\T \in \TT_u$ we denote by
$X_u(j,\C,\T)$ the minimum number of influenced nodes one can attain in $G[W(u)]$ by immunizing at most $j$ nodes in $W(u)$, where the states and the thresholds of nodes in $W_u$ are given by  $\C$ and $\T$.
\end{definition}
Considering that the root $r$ of a nice tree decomposition has  $W_r = \emptyset$, we have that the solution of the IDM instance $\langle G, \l\rangle$  can  be obtained by computing  $X_r(\l,\emptyset,\emptyset).$

\begin{claim} \label{claimTW}
For each  $u\in T$, the computation of $X_u(j,\C,\T)$,  for each    $j\in\{0,\ldots, \ell\}$,  state configuration  $\C\in \CC_u$, and  threshold configuration $\T \in \TT_u$  comprises $O(\l3^\tw \mu^{\tw})$ values,  where $\mu=\min\{k, \Delta\}$, each of which can be computed recursively in time $O(2^\tw+\l)$.
\end{claim}

\begin{proof}
We show now how use a bottom--up strategy to compute all the values of $X_u(j,\C,\T)$,  for each  $u\in T$,   $j=0,\ldots,\l$,  state configuration  $\C\in \CC_u$, and  threshold configuration $\T \in \TT_u$. By Definition \ref{eq_Xu}, we know that such values are $O(\l3^\tw \mu^{\tw})$,  where $\mu=\min\{k, \Delta\}$.
\\
For each leaf $u \in T$ and for each $j=0,\ldots,\l$ we have $ X_u(j,\emptyset,\emptyset)= 0.$
\\
For any internal node $u$, we show how to compute each values $X_u(j,\C,\T)$,  for each $j=0,\ldots, \l$,     $\C \in \CC_u$,   and    $\T \in \TT_u$ in time {$O(2^\tw+\l)$}.

We have three cases to consider according to the type of $u$ (cf. Definition \ref{nice}):

\begin{description}
	\item[1) $u$ is an introduce node:] In this case $u$ has exactly one child $u'$ and we have that $W_u=W_{u'} \cup \{v\}$ for some node $v \notin W_{u'}$. 
	For a given  node $u \in V(T)$ (introducing a node $v \in V$) and state configuration $\C$, we denote by $\XWG{u}{\C}$ the set of influenced nodes (according to the configuration $\C$)  that belongs to $W_u \cap{\Gamma_G(v)}$ . Given a threshold configuration $\T$ associated to a set of nodes $W$, and a set of nodes $S\subseteq W$ we denote by $\Tminus{\T}{S}$ the configuration obtained starting from $\T$ and decreasing by one the threshold of each node in $S.$ 
	 {In the following we assume w.l.o.g. that  the element indexed by $v$ is the last element of the vectors $\C$ and $\T$.}   
	We have that for each $j=0,\ldots, \l$, each $\C \in \CC_u$  and each $\T \in \TT_u.$
		\begin{equation}
		X_u(j,\C{=}[\C',c],  \T{=}[\T',t])=
		\begin{cases} 
{ min_{S\subseteq \XWG{u}{\C}, |S|=t} \big(X_{u'}(j,\C',\Tminus{\T'}{\XWG{u}{\C}{-}S}) \big){+} 1},
\\
\qquad \qquad \mbox{if } c=\xx \mbox{ AND } t \leq |\XWG{u}{\C}|   \\
  X_{u'}(j-1,\C',\T'),\\
\qquad \qquad  \mbox{if } c=\yy  \mbox { AND } j>1 \\
	X_{u'}(j,\C',\T'), \\
\qquad \qquad	\mbox{if } c=\zzz  \mbox { AND }  t > |\XWG{u}{\C}|  \\
+\infty, \qquad
 \mbox{otherwise. }\label{eq:introduce}
\end{cases} 
\end{equation} 
It is worth to observe that the size of $\XWG{u}{C}$ is bounded by $\tw$ and for this reason the above value can be computed in time $O(2^\tw)$

\item[2) $u$ is a  forget node:] In this case $u$ has exactly one child $u'$ and we have that $W_{u'}=W_{u} \cup \{v\}$ for some node $v \notin W_u$. We have for each $j=0,\ldots, \l$, each $\C \in \CC_u$,  and each $\T \in \TT_u$
		\begin{equation}\label{eq:forget}
		X_u(j,\C,  \T)=min_{c\in\{\xx,\yy,\zzz\}}\{X_{u'}(j,\C'=[\C,c],\T'=[\T,\max\{0,t(v)-|\XWG{u}{\C}|\}] )\}
	\end{equation}

\item[3) $u$ is a  join node:] In this case $u$ has exactly two child $u_1$, $u_2$ such that $W_{u} = W_{u_1} = W_{u_2}$.
We have for each $j=0,\ldots, \l$, each $\C \in \CC_u$,   and each $\T \in \TT_u$
		\begin{equation}\label{eq:join}
		X_u(j,\C,  \T)=min_{0\leq a\leq j-I(\C)}\{X_{u_1}(a+I(\C),\C,\T)+\{X_{u_2}(j-a,\C,\T)\},
	\end{equation}
	where $I(\C)$ denotes the number of immunized nodes in the configuration state $\C.$ 
\end{description}

By induction on the tree, we can prove that the recursive formula presented in  (\ref{eq:introduce})-(\ref{eq:join}) coincides with the  definition of $X_u(\cdot, \cdot, \cdot)$; hence, the algorithm is correct. 
\end{proof}

Hence, using [\citealp{Treewidth_book}, Lemma 18], we have that the desired value $X_r(\l,\emptyset,\emptyset))$ can be computed in time
$O(\tw|V|(2^\tw+\l)\l3^\tw \mu^{\tw}).$
Standard backtracking techniques can be used to compute the optimal set $X$ and $Y(X)$ in the same time.
\\
%
%
As a consequence we have that {\sc IDM} is FPT  with respect to    \tw\  and  $\Delta$ or $k$.
\begin{theorem}
 {\sc IDM}  is solvable in  time $O(\tw|V|(2^\tw+\l)\l3^\tw \mu^{\tw})$, where $\mu=\min\{k, \Delta\}$.
\end{theorem}

\subsection{Graphs of bounded neighborhood diversity}
We present FPT algorithms for  \IIB\    with respect to both the pairs  $(k,\nd)$ and $(\ell,\nd)$.

Let  $\{V_1,V_2, \ldots, V_\nd\}$ be the type partition of $G$. 
%
Below, we assume that the nodes of each $V_i=\{v_{i,1},\ldots,v_{i,|V_i|}\}$ are sorted 
in  non-decreasing order of thresholds, e.g. $t(v_{i,j})\leq t(v_{i,j+1})$.

\medskip

\noindent
{\large \textbf{Parameters $\nd$ and $\k$.}}
We consider all the  
$\nd$-ples $(f_1,\ldots,f_\nd)$ such that $\sum_{i=1}^{\nd} f_i \leq k$. 
For each one, we construct a candidate set as detailed in Algorithm IIB-k below.
\smallskip
  \begin{algorithm}[tb!]
\SetCommentSty{footnotesize}
\SetKwInput{KwData}{Input}
\SetKwInput{KwResult}{Output}
\DontPrintSemicolon
\caption{ \ \    IIB-k($G, \k, \l$) \label{alg1-1}}
\KwData {A graph $G=(V,E,t)$,  integers $\k, \l$ and  a type partition $V_1, \ldots, V_\nd$ of $G$.}
\setcounter{AlgoLine}{0}
\ForEach {$\f=1, \ldots, \k$}{
\ForEach { $\bbf=(f_1,f_2,\ldots,f_\nd)$ such that $\sum_{i=1}^{\nd} f_i =\f$}{
  \lForEach   {$i\in[\nd]$} {let $X_i=\{v_{i,1},\ldots,v_{i,f_i}\}\subseteq V_i$}     
  Set $X = \bigcup_{i=1}^{\nd} X_i$\\
  \lIf{ $|Y(X)| \leq \l$ }{
                 \textbf{return} {\sc yes}
								}	
	 }
}
\textbf{return} {\sc no}
\end{algorithm}

\begin{theorem}\label{teorema-nd-k}
Algorithm IIB-k  solves  \IIB\    
  in time  $O(n^2\,   2^{k+nd-1} )$
  \end{theorem}
  
\begin{proof}
We first show  Algorithm IIB-k outputs {\sc yes} iff there exists  $ X$ satisfying (\ref{equiv}).

If the output is {\sc yes} then trivially the current set $X$    has $X\leq k$ and  $|Y(X)|\leq\ell$.

Let now $\tilde X$ be  a minimal set  satisfying (\ref{equiv}), that is,
$\tilde X=\A_{G[\tilde X]}$, $|X|\leq k$, and $|Y(X)|\leq \ell$.
Let $\tilde X_i=\tilde X \cap V_i$ for each  $i\in [\nd]$. 
Consider the iteration of the algorithm when  $\bbf=(f_1,f_2,\ldots,f_\nd)$ with $f_i=|\tilde X_i|$, for  $i\in [\nd]$.
The algorithm selects a set 
 $X = \bigcup_{i=1}^{\nd} X_i$ such that
$|X_{i}|=f_i$ and $t(v)\leq t(w)$ for each $v\in X_i$ and $w\in V_i-X_i$, for each $i\in [\nd]$. We show that the algorithm outputs {\sc yes} on $X$.
\\
Fix any $i\in[\nd]$.  Knowing that  $|\tilde X_i|=|X_{i}|=f_i$, we have that if $\tilde X_i \neq X_{i}$, 
 then there exists   
$u\in \tilde X_i-X_{i}$ and  $v\in X_{i}-\tilde X_i$ such that $t(v) \leq t(u)$.
W.l.o.g assume that $u$ is the node with maximum threshold in $\tilde X_i-X_{i}$.
Since $\tilde X= \Active_{G[\tilde X]}$,
we have that $u$ has at least $t(u)$ neighbors in $\tilde X$.
Furthermore, since $v,u \in V_i$ we have that $u$ and $v$ have the same neighbors. Hence, $v$ has at least $t(u)\geq t(v)$ neighbors in $\tilde X$. As a consequence,  since $v\notin \tilde X$ we have  $v \in Y(\tilde X)$. 
Consider $\tilde X'=\tilde X -\{u\} \cup \{v\}$. By (i) in Proposition \ref{prop-change} (see Appendix)
we have that $\tilde X'= \Active_{G[\tilde X']}$ with
$|\tilde X'|=|\tilde X|$ and $|Y(\tilde X')|=|Y(\tilde X)|$. 

Hence, trading each node in $\tilde X_i-X_{i}$ for one in $X_i-\tilde X_i$, for each $i$ such that $\tilde X_i \neq X_i$, we can prove that 
$|Y(X)|=|Y(\tilde X)|\leq \l$. 
Therefore, the algorithm  returns {\sc yes}.

We now evaluate the running time.
Fix $\f\in [\k]$, for each  $(f_1,\ldots,f_\nd)$ with
$\sum_{i=1}^{\nd} f_i = f$, one needs 
time $O(f)$  to get $X $ and $O(n^2)$  to get $Y(X)$, moreover the number of all  possible such \nd-ple is 
 $\binom{f + \nd -1}{f}$.
 Summing on all $f$ we get $ \sum_{f\in [k]}\binom{f + \nd -1}{f}<2^{k+\nd-1}$ and the theorem holds.
\end{proof}

\medskip

\noindent
{\large \textbf{Parameters $\nd$ and $\l$.}} An idea similar to that in  Algorithm 1 can be used to prove \IIB\, is FPT with respect to \nd\, and $\ell$.

\smallskip

\begin{algorithm}[tb!]
\SetCommentSty{footnotesize}
\SetKwInput{KwData}{Input}
\SetKwInput{KwResult}{Output}
\DontPrintSemicolon
\caption{ \ \    IIB-$\l$($G, \k, \l$) \label{alg2-1}}
\KwData {A graph $G=(V,E,t)$,  integers $\k, \l$ and  a type partition $V_1, \ldots, V_\nd$ of $G$.}
\setcounter{AlgoLine}{0}
\ForEach {$\h=1, \ldots, \l$}{
\ForEach { $\bh=(\h_1,\h_2,\ldots,\h_\nd)$ such that $\sum_{i=1}^{\nd} \h_i =\h$}{
\lForEach  {$i\in[\nd]$} {let $Y_i=\{v_{i,1},\ldots,v_{i,h_i}\}\subseteq V_i$}  
 Set $\Y = \bigcup_{i=1}^{\nd} \Y_i$\\
 \lIf{$|\Active_{G,\Y}| \leq \k$}{\textbf{return} {\sc yes}}
}
}
\textbf{return} {\sc no}
\end{algorithm}

\begin{prop}\label{prop-change}
Fix $i\in [\nd]$.
\begin{itemize}
\item[(i)] Let $X=\A_{G[X]}$  and $Y=Y(X)$ be its immunizing set. Set 
 $u_{max}=\argmax_{u\in X \cap V_i}t(u)$.
If there exists $v \in Y \cap V_i$ such that $t(v) \leq  t(u_{max})$ then $X'=X - \{u_{max}\} \cup \{v\}$ satisfies $X'= \Active_{G[X']}$  and $|Y(X')|=|Y|$.
\item[(ii)]  Let   $Y$ be an  immunizing set. Set $v_{max}=\argmax_{v\in Y \cap V_i}t(v)$. If there exists $u \in \Active_{G,Y} \cap V_i$ 
such that $t(u) \leq  t(v_{max})$ then setting $Y'=Y - \{v_{max}\} \cup \{u\}$  it holds $|\Active_{G,Y'}|\leq|\Active_{G,Y}|$.
\end{itemize}
\end{prop}
\begin{proof}
Let us prove (i).
Consider $X'=X -\{u_{max}\} \cup \{v\}$ and the diffusion process in 
$G[X']$.
We have that $v$ is influenced at a round which is at most equal to that  in which $u_{max}$ is influenced during the diffusion process in $G[X]$ 
(recall $t(v) \leq t(u_{max})$ and that $v$ and $u_{max}$ have the same neighbors). 
Furthermore, since all the neighbors of $v$ and $u_{max}$ have the same number of neighbors in $X'$ as in $X$  we have that all the nodes in $X'$ are influenced, that is    $X'= \Active_{G[X']}$, and  $u_{max} \in Y(X')$.  
This allows to say that 
$|Y(X')|=|Y|$. 
\\
Let us prove now (ii).
If we consider the diffusion process in $G[V-Y']$  we have that no node outside $\Active_{G,Y}-\{u\}$, except eventually for node $v_{max}$, can be 
influenced. 
Hence, $\Active_{G,Y'} \subseteq \Active_{G,Y}-\{u\}\cup \{v_{max}\}$.
\end{proof}

\begin{theorem}\label{teorema-nd-l}
Algorithm IIB-$\l$  solves  \IIB\    
  in time  $O(n^2\,   \ 2^{\ell+\nd-1} )$
\end{theorem}

\begin{proof}
Given $\h \leq \l$,  Algorithm IIB-$\ell$($G, \k, \l$)  considers all the possible 
$\nd$-ples $(\h_1,\h_2,\ldots,\h_\nd)$ with $\sum_{i=1}^{\nd} \h_i =\h$;
 for each   $\bh=(\h_1,\h_2,\ldots,\h_\nd)$ we construct 
the set $\Y= \bigcup_{i=1}^{\nd} \Y_i$ 
where $\Y_i$ consists of  the first (e.g.    with the smallest thresholds) $\h_i$ nodes in $V_i$.
We then consider the diffusion process  in $G$ and
     the 
set $\Active_{G,Y}$  of influenced nodes  when the elements of  $Y$ are immunized.
If $|\Active_{G,Y}| \leq \k$ then we answer {\sc yes}.
In case no  $\bh$ gives 
a set $Y$ such that 
 $|\Active_{G,Y}| \leq k$, we answer {\sc no}.
 
If Algorithm IIB-nd-$\ell$ returns {\sc yes} then  the
 set $Y$  constructed by algorithm IIB-$\ell$ has size at most $\l$ and we know that $|\Active_{G,Y}| \leq \k$.

Assume now that there exists $\tilde Y$ such that $|\tilde Y|=h\leq \l$  and  $|\Active_{G,\tilde Y}|\leq k$.   Assume w.l.o.g. that no smaller solution exists, that is, for any $Y$ such that $|\Active_{G,Y}|\leq k$ it holds $|Y|\geq h$. 

Define $\tilde Y_i=Y(\tilde X) \cap V_i$ and let $|\tilde Y_i|=h_i$, for $i\in [\nd]$. 
Clearly,
 $\sum_{i=1}^{\nd}h_i=h$.
Consider  the \nd-ple $\bh=(h_1,h_2,\ldots,h_\nd)$ and the set 
$Y= \bigcup_{i=1}^{\nd} Y_{i}$   constructed at line 4 of algorithm IIB-nd-$\ell$.
Recall that  $|Y_{i}|=\h_i$  and $t(v)\leq t(w)$ for each $v\in Y_i$ and $w\in V_i-Y_i$.
\\
Since $|\tilde Y_i|=|Y_{i}|=h_i$, we have that if $\tilde Y_i  \neq Y_{i}$, 
for some $i$, then there are
$v\in \tilde Y_i -Y_{i}$ and  $u\in Y_{i}-\tilde Y_i$ such that
$t(u) \leq t(v)$.
W.l.o.g select $u$ as the node with minimum threshold in
$Y_{i}-\tilde Y_i$ and $v$  as the node with maximum threshold in
$\tilde Y_i -Y_{i}$.
By the fact that  $v \in  \tilde Y$ and $\tilde Y$ is minimal, we know that $v$ must have at least $t(v)$ neighbors in $\Active_{G,\tilde Y}$ (otherwise, $\tilde Y-\{v\}$ would be a smaller solution).
Furthermore, since $v,u \in V_i$ we have that they have the same neighbors. As a consequence, also  $u$ has at least $t(v)\geq t(u)$ neighbors in $\Active_{G,\tilde Y}$.
Knowing that $u\not \in \tilde Y$, we have that $u \in \Active_{G,\tilde Y}$.
Set $Y'=\tilde Y-\{v\} \cup \{u\}$.
By (ii) in Proposition \ref{prop-change}
we have that  $ \Active_{G,Y'}$
 satisfies   $\Active_{G,Y'} \leq \Active_{G,\tilde Y} \leq k$. 
Hence, $Y'$ is also a solution.

Starting from  $Y'$, we then can repeat the above reasoning until we get $Y^r=Y$, the immunizing set considered in the algorithm for the tuple $\bh$. Hence,  
$|\Active_{G,Y}|\leq k$.

Now we evaluate the running time of the algorithm.
For each fixed $\h\in [\ell]$, the number of all the possible 
$\nd$-ples $(h_1,h_2,\ldots,h_\nd)$ such that 
$\sum_{i=1}^{\nd} h_i = h$ is 
$\binom{h + \nd -1}{h} \leq \binom{\ell + \nd -1}{h}.$
Noticing that 
 for each choice of  $(h_1,\ldots,h_\nd)$, one needs time  $O(h)$ 
 to construct $Y$ and $O(n^2)$ to obtain
$\Active_{G,Y}$ and that 
$$\sum_{h\in[\ell]}\binom{\ell + \nd -1}{h}<2^{\ell + \nd -1},$$
the desired result follows.
\end{proof}

\section{Conclusion}
{We introduced the influence immunization problem on networks under the threshold model and  analyzed its  parameterized complexity. We considered several parameters and showed that the problem remains intractable with respect to each  one. We have  also shown that  for some pairs (e.g., ($\zeta$, $\l$) and ($\Delta$, $\l$)) the problem remains intractable. 
\\
On the positive side, the problem was shown to be  FPT  for some other pairs:  $(k,\ell)$, $(k, \zeta)$, $(k,\tw), (\Delta,\tw), (k,\nd)$, and $(\ell,\nd)$. 
\\
It would be interesting to asses the parameterized complexity of \IIB\, with  respect to the remaining  pairs of parameters; in particular  with respect to  $k$ and $\Delta$.
}


\begin{thebibliography}{99}

\bibitem{ALM+}F.N. Abu-Khzam, S. Li, C. Markarian, F. Meyer auf der Heide, P. Podlipyan.
Modular-Width: An Auxiliary Parameter for Parameterized Parallel Complexity. Proc. of Frontiers in Algorithmics. (FAW 2017), LNCS, v. 10336. Springer,  (2017).

\bibitem{barabasi}
R. Albert, H. Jeong, A.-L. Barab\'asi.
Error and attack tolerance of complex networks,
Nature, vol. 404, 378-382, (2000).

\bibitem{CBFGR}R. Belmonte, F.V. Fomin,  P.A. Golovach, M.S. Ramanujan. 
Metric Dimension of Bounded Width Graphs. Proc. Mathematical Foundations of Computer Science  (MFCS '15), LNCS vol 923, Springer, (2015).

\bibitem{Ben-Zwi}
O. Ben-Zwi, D. Hermelin, D. Lokshtanov, I. Newman.
Treewidth governs the complexity of target set selection.
Discrete Optimization, vol. 8(1), 87--96, ISSN 1572-5286, https://doi.org/10.1016/j.disopt.2010.09.007,  (2011).

\bibitem{Kempe-vaccino}
P. Chen, M. David, D. Kempe.
Better Vaccination Strategies for Better People, 
 Proceedings 11th ACM Conference on Electronic Commerce (EC-2010), Cambridge, Massachusetts, USA, June 7-11, (2010).
 
\bibitem{CGMRV} G. Cordasco, L. Gargano, M. Mecchia, A. A. Rescigno, U. Vaccaro, 
Discovering Small Target Sets in Social Networks: A Fast and Effective Algorithm, Algorithmica,  80(6),  1804-1833, (2018).

\bibitem{CGRV}	G. Cordasco, L. Gargano, A. A. Rescigno, U. Vaccaro. 
Evangelism in social networks: Algorithms and complexity. In Networks 71(4): 346--357, (2018).

\bibitem{itp} 	G. Cordasco, L. Gargano, A. A. Rescigno.
Iterated Type Partitions, Proceedings IWOCA 2020, 195--210, (2020).

\bibitem{asonam}
G. Cordasco, L. Gargano, A. A. Rescigno,
Influence propagation over large scale social networks,
Proceedings of the 2015 IEEE/ACM International Conference on Advances in Social Networks Analysis and Mining, ASONAM 2015, 1531-1538.

\bibitem{CDP} D. Coudert, G. Ducoffe,  A. Popa. 
 Fully polynomial FPT algorithms for some classes of bounded clique-width graphs. In Proceedings of the Twenty-Ninth Annual ACM-SIAM Symposium on Discrete Algorithms  (SODA '18), 2765--2784, (2018).
 
 \bibitem{CFKLMPPS15}
M. Cygan, F.V. Fomin, L. Kowalik, D. Lokshtanov, D. Marx, M. Pilipczuk, M. Pilipczuk, and S. Saurabh.
Parameterized Algorithms. Springer, doi:10.1007/978-3-319-21275-3, (2015).
 
\bibitem{DF}
 {R.G. Downey and M.R. Fellows}. 
{ {Parameterized Complexity}}, {Springer}, {(2012)}.

\bibitem{DKT16}
P. Dvor\'ak, D. Knop, and T. Toufar. 
  Target Set Selection in Dense Graph Classes.  
Proc. 29th International Symposium on Algorithms and Computation  (ISAAC 2018), 10.4230/LIPIcs.ISAAC.2018.18, (2018).

\bibitem{ER19}
S. Ehard, D. Rautenbach.
Vaccinate your trees!
Theoretical Computer Science, vol. 772,  46--57, ISSN 0304-3975, https://doi.org/10.1016/j.tcs.2018.11.018, (2019)

\bibitem{feige}
U. Feige, R. Krauthgamer, K. Nissim.
On cutting a few vertices from a graph,
Discrete Applied Mathematics, 127, 643 – 649,  2003. 

\bibitem{cuttingedge}
E. B. Khalil, B. Dilkina, L. Song.
CuttingEdge: Influence minimization in networks.
Workshop on Frontiers of Network Analysis: Methods, Models, and Applications at NIPS, (2013).

\bibitem{Treewidth_book}
T. Kloks
Treewidth Computations and Approximations
Lecture Notes in Computer Science vol. 842, Springer-Verlag Berlin Heidelberg, ISSN 0302-9743,  10.1007/BFb0045375, (1994).

\bibitem{FGKKK}
J. Fiala, T. Gavenciak, D. Knop, M. Koutecky, J. Kratochv\'il.  
Fixed parameter complexity of distance constrained labeling and uniform channel assignment problems.
In {arXiv:1507.00640}, (2015).	

\bibitem{Fomin}
F. V. Fomin, P. A. Golovach, J. H. Korhonen.
On the Parameterized Complexity of Cutting a Few Vertices from a Graph.
International Symposium on Mathematical Foundations of Computer Science
(MFCS 2013), LNCS, vol, 8087, pp. 421-432 (2013).

\bibitem{G}
 R. Ganian.
Using neighborhood diversity to solve hard problems.  
\texttt{arXiv:1201.3091}, (2012).

\bibitem{GR}
 {L. Gargano, A.A. Rescigno}.
 {Complexity of conflict-free colorings of graphs}.
 { Theoretical Computer Science }, 566,  {39--49}, (2015).


\bibitem{GJ79}
{M. Garey, D. Johnson}.
Computers and Intractability: A Guide to the Theory of NP-Completeness, Freeman, San Francisco, (1979).

\bibitem{GKK18}
T. Gavenciak, D. Knop and M. Kouteck{\'{y}}.
Integer Programming in Parameterized Complexity: Three Miniatures
In Proc. of 13th Intern. Symp. on Parameterized and Exact Computation, {IPEC} 2018, 10.4230/LIPIcs.IPEC.2018.21, (2018).

\bibitem{granovetter}
M. Granovetter. 
Threshold models of collective behaviors. 
The American
Journal of Sociology, 83(6), 1420–1443, (1978).

\bibitem{Kempe-cut}
A. Hayrapetyan, D. Kempe, Ma. P. Svitkina. 
Unbalanced Graph Cuts.
Proc. European Symposium on Algorithms
(ESA 2005), LNCS 3669, pp 191-202|, (2005).

\bibitem{bodlander}
T. Hanaka, H. L.Bodlaender, T. C.van der Zanden, H. Ono, 
On the maximum weight minimal separator.
Theoretical Computer Science
vol. 796, pp. 294-308, (2019).

\bibitem{Kempe}
 D. Kempe, J. Kleinberg, E. Tardos.  
Maximizing the spread of influence through a social network. 
 In Proc. of the 9th ACM SIGKDD int. conf. on knowledge
discovery and data mining, Washington, USA, 137--146, (2003).

\bibitem{Kimuraetal}
M. Kimura,  K. Saito, H.  Motoda.
Blocking links to minimize contamination
spread in a social network, ACM Trans. on Knowledge Discovery from Data,
3(2), 9, (2009).

\bibitem{subgraphs} 
C. Komusiewicz, M. Sorge.
Finding Dense Subgraphs of Sparse Graphs.
Parameterized and Exact Computation,  LNCS, vol. 7535, (2012).

\bibitem{Kn19}
D. Knop, M. Kouteck{\'{y}},  T. Masar{\'{\i}}k,  T. Toufar.
Simplified Algorithmic Metatheorems Beyond {MSO:} Treewidth and Neighborhood  Diversity.
Logical Methods in Computer Science 15 (4),  (2019).

\bibitem{L}
M. Lampis. Algorithmic meta-theorems for restrictions of treewidth. Algorithmica 64, 19--37, (2012). 
\bibitem{book} F. Menczer, S. Fortunato,  C. A. Davis, A First Course in Network Science, Cambridge University Press; 1st edition (2020).
\bibitem{newman} M. E. J. Newman, S. Forrest, J. Balthrop.
Email networks and the spread of computer viruses. Physical Review E, vol. 66, (2002).

\bibitem{N}
R. Niedermeier.  Invitation to Fixed-Parameter Algorithms. Oxford University Press, (2006).

\bibitem{R86}
N. Robertson and P.D. Seymour. 
Graph minors. II. Algorithmic aspects of tree-width. 
Journal of Algorithms, vol. 7(3), 309--322, (1986).
\end{thebibliography}
\end{document}